\documentclass[aps,amsfonts,pre,twocolumn,superscriptaddress,showpacs]{revtex4}
\usepackage{epsfig,amsmath,amssymb,bm,epsf,graphics,psfrag,verbatim,subfigure,mathrsfs}
\usepackage[usenames]{color}

\def\Prob{\mathcal{P}}
\def\Disp{u}
\def\freq{\omega}
\def\freqst{\freq^{*}}
\def\lst{l^{*}}

\def\EE{U}
\def\RefrPosil{\mathbf{R}_{\ell 0}}
\def\TargPosil{\mathbf{R}_{\ell}}
\def\RefrPosilp{\mathbf{R}_{\ell' 0}}
\def\TargPosilp{\mathbf{R}_{\ell'}}
\def\ul{\mathbf{u}_{\ell}}
\def\ulp{\mathbf{u}_{\ell'}}
\def\uv{\mathbf{e}}
\def\ub{\mathbf{u}_{b}}
\def\DM{\mathbf{D}}
\def\vu{\mathbf{u}}
\def\vq{\mathbf{q}}
\def\fcNN{k}
\def\fcNNN{\kappa}
\def\vB{\mathbf{B}}
\def\qst{q^{*}}
\def\fcNNNm{\kappa_{m}}
\def\fcNNNs{\kappa_{s}}
\def\PPV{\mathbf{V}}
\def\GF{\mathbf{G}}
\def\TM{\mathbf{T}}
\def\fFunc{f}
\def\DOS{\rho}

\def\Diag{\textrm{Diag}}
\def\tGF{\tilde{\mathbf{G}}}
\def\ulone{\mathbf{u}_{\ell_1}}
\def\ultwo{\mathbf{u}_{\ell_2}}
\def\bv{\mathbf{b}}
\def\rv{\mathbf{r}}
\newcommand{\lv}{{\mathbf{l}}}
\newcommand\bb{\mathcal{B}}
\newcommand\Aa{\mathcal{A}}

\begin{document}
\title{Coherent potential approximation of random nearly isostatic kagome lattice}
\author{Xiaoming Mao}
\affiliation{Department of Physics and Astronomy, University of Pennsylvania, Philadelphia, PA 19104, USA }
\author{T. C. Lubensky}
\affiliation{Department of Physics and Astronomy, University of Pennsylvania, Philadelphia, PA 19104, USA }

\date{\today}

\begin{abstract}
The kagome lattice has coordination number $4$, and it is mechanically isostatic when nearest neighbor ($NN$) sites are connected by central force springs. A lattice of $N$ sites has $O(\sqrt{N})$ zero-frequency floppy modes that convert to finite-frequency anomalous modes when next-nearest-neighbor ($NNN$) springs are added.  We use the coherent potential approximation (CPA) to study the mode structure and mechanical properties of the kagome lattice in which $NNN$ springs with spring constant $\kappa$ are added with probability $\Prob= \Delta z/4$, where $\Delta z= z-4$ and $z$ is the average coordination number.  The effective medium static $NNN$ spring constant $\kappa_m$ scales as $\Prob^2$ for $\Prob \ll \kappa$ and as $\Prob$ for $\Prob \gg \kappa$, yielding a frequency scale $\omega^* \sim \Delta z$ and a length scale $l^*\sim (\Delta z)^{-1}$. To a very good approximation at at small nonzero frequency, $\kappa_m(\Prob,\omega)/\kappa_m(\Prob,0)$ is a scaling function of $\omega/\omega^*$. The Ioffe-Regel limit beyond which plane-wave states becomes ill-define is reached at a frequency of order $\omega^*$.
\end{abstract}


\maketitle

\section{Introduction}
\begin{figure}
\centerline{\includegraphics[width=.4\textwidth]{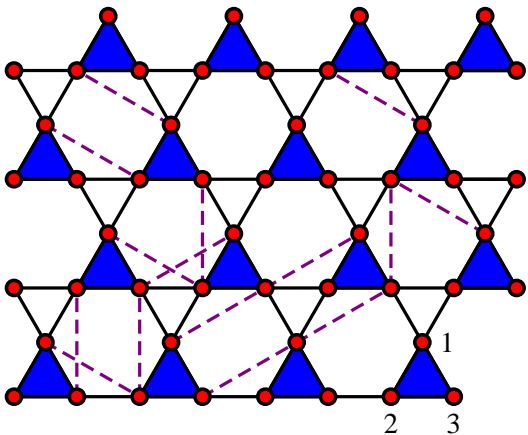}}
\caption{The kagome lattice with random additional $NNN$ bonds denoted by purple dashed lines.  The unit cell triangle is marked with filled triangles.  Particle $1$, $2$, $3$ in each unit cell are marked in the bottom right unit cell.}
\label{FIG:KagoRand}
\end{figure}

Understanding the nature of mechanical stability, how it arises, and how it can be controlled is important to fields ranging from civil engineering to biology.  James Clerk Maxwell was among the first to address this problem with mathematical rigor in his 1864 paper \cite{Maxwell1864} in which he established the conditions for the mechanical stability of a frame of points connected by lines or struts of fixed length.  His approach, which can easily be generalized to treat particles interacting via central-force potentials or bonds between particles interacting via an angular potential, has since been used to study network glasses \cite{Phillips1985,Thorpe1983,He1985}, randomly packed spheres near jamming \cite{Wyart2005,Wyart2005a}, networks of semi-flexible polymers \cite{Heussinger2006,Huisman2010}, regular \cite{Souslov2009} and random \cite{Mao2010} periodic lattices, and a host of engineering problems \cite{Kapko2009,Calladine1978}.  
In this paper, we study the mode structure of a particular lattice - the kagome lattice shown in Fig.~\ref{FIG:KagoRand} - in which nearest neighbor ($NN$) sites are connected by harmonic central-force springs of spring constant $\fcNN$ and next-nearest-neighbor ($NNN$) sites are randomly connected by similar springs but with a different spring constant $\fcNNN$.  In the absence of $NNN$ springs, finite realizations of this lattice are just on the verge of mechanical stability in the Maxwell sense.  The addition of $NNN$ springs renders the system stable.  When these springs are added uniformly with a spring constant $\kappa$ that is allowed to approach zero continuously \cite{Souslov2009,Souslov2010}, the result is a kind of critical point at $\kappa=0$ with a rich mode structure and diverging length and vanishing frequency scales as $\kappa \rightarrow 0$ that have direct counterparts in the properties of randomly packed spheres near jamming \cite{Liu1998}.  Here we use the coherent potential approximation or CPA \cite{Soven1969,Feng1985} (otherwise know as the effective medium approximation) to explore in detail the static and dynamical mechanical properties and mode structure of the kagome lattice as a function of the probability $\Prob$ that $NNN$ sites are connected by springs.  We find that the effective medium $NNN$ spring constant $\fcNNNm(\Prob,\freq)$ scales as $\Prob^2$ for small $\Prob$, and it gives rise to length and frequency scales $\lst\sim\Prob^{-1}$ and $\freqst\sim\Prob$.  Beyond a characteristic frequency $\freqst_D\sim\Prob$, the imaginary part of $\fcNNNm$ rises quickly and phonons scatter strongly.  The Ioffe-Regel limit beyond which certain plane wave states become ill-defined is reached at a frequency slightly larger than $\freqst_D$.

\subsection{Mechanical Stability and Floppy Modes}

Maxwell \cite{Maxwell1864} argued that each constraint in a system of $N$ points in $d$-dimensions reduces the number, $dN$, of zero frequency modes of free points by one.  Thus, the number of zero frequency modes in a system with $N_c$ constraints is $N_0=d N - N_c$.  Of these, $d(d+1)/2$ are those of rigid translations and rotations, leaving $N_f = dN - N_c - d(d+1)/2$ internal zero frequency modes, which are usually refereed to as floppy modes \cite{Thorpe1983} or mechanisms in the engineering literature~\cite{Calladine1978}, when $dN - N_c - d(d+1)/2\geq 0$.  Mechanical stability requires that there be no floppy modes, or equivalently that $dN - N_c - d(d+1)/2\leq 0$.  Though Maxwell considered general frames of points, his analysis can immediately be applied to the determination of mechanical stability of extended crystals and glasses - as was first done in the context of network glasses, which are unstable in the absence of bending forces, by Phillips and Thorpe \cite{Phillips1985,Thorpe2000} - in which points are replaced by particles or atoms and struts are replaced by inter-particle potentials.

If each particle in an extended system is on average connected by central-force potential creating bonds to $z$ other particles, then, because each bond is shared by two particles, the system experiences $zN/2$ constraints if bond lengths are restricted to be at their equilibrium length and there are no redundant bonds \cite{Jacobs1995}, and it has $N_f=dN - \frac{1}{2}z N - \frac{1}{2} d(d+1)$ floppy modes.  Thus, in the limit of large $N$, a central-force system is mechanically stable if $z>z_c = 2d$.  In an infinitely extended system, $z$ is independent of $N$.  In a finite system cut from an infinite one, particles at the boundary have fewer neighbors than those in the interior, and $z$ (which is an average quantity) is of order $N^{-1/d}$ smaller than it would be in the ideal infinite system.  Systems with $z=z_c$ are called isostatic.  A finite piece of $N$ particles cut from an infinite isostatic has a surface with of order $N^{(d-1)/d}$ cut bonds, and thus that many fewer constraints and that many more floppy modes.  
These modes generally remain floppy even if they have finite amplitudes if there are free boundary conditions.  Because of their geometry with straight segments extending throughout their interior, the square, the kagome, and related lattice of $N$ sites with periodic rather than free boundary conditions have $N^{(d-1)/d}$ infinitesimal floppy modes - modes that have zero energy for infinitesimal displacements but finite energy for large displacements \cite{Calladine1978}.  The simplest example of an infinitesimal floppy modes is that of a mass connected to two identical co-linear springs that are rigidly attached to walls on either side of the mass.  If the springs are not under tension, infinitesimal displacements of the particle perpendicular to the spring cost no energy but finite displacements do.  Thus the harmonic kagome lattice has $\sqrt{N}$ floppy modes.

The recently most explored random isostatic system is perhaps the system of randomly close-packed spheres at the jamming transition \cite{Liu1998} in the system of frictionless soft spheres with one-sided repulsion.  This transition occurs when the volume faction $\phi$ of spheres just exceeds the critical value $\phi_c$ at which they form just enough contacts to support a compressional load.  
Thus at $\phi_c$, the bulk modulus $B$ jumps discontinuously to a nonzero value, the number of neighbors per particle grows as $\Delta z = z-z_c =(\phi-\phi_c)^{1/2}$, and the shear modulus grows continuously from zero as $G \sim (\phi-\phi_c)^{1/2} \sim (\Delta z)^{1}$.  Associated with this transition are a diverging length scale $l^* \sim (\Delta z)^{-1}$ and vanishing frequency scale $\omega^* \sim \Delta z$ whose behavior follows from quite general ``cutting" arguments \cite{Wyart2005}.

\subsection{Periodic Isostatic Lattices}

Isostaticity is not limited to random systems. There are a number of periodic isostatic lattices with $z_{NN}= z_c$ $NN$ bonds per particle.  These include the hypercubic lattice in $d$ dimensions and the kagome lattice in two-dimensions and its generalization to higher dimensions with $NN$ bonds occupied by springs of spring constant $k$, all with $z=2d$.  They can be moved off isostaticity by introducing springs of spring constant $\kappa$ on $NNN$ bonds either homogeneously or randomly.

In homogeneous systems, the isostatic limit is continuously approached by allowing $\kappa \rightarrow 0$.  In the square-lattice version of this model \cite{Souslov2009}, the bulk modulus is nonzero at $\kappa = 0$, but the shear modulus vanishes as $\kappa$ in this limit. In the kagome version, both the shear and the bulk moduli are nonzero at $\kappa = 0$.  In both models, as in the jamming problem, there is a divergent length scale, $l^* \sim \kappa^{-1/2}$ and a vanishing frequency scale $\omega^*\sim \kappa^{1/2}$.

The Coherent Potential, or equivalently the effective medium, Approximation (CPA) \cite{Lax1951,Taylor1967,Soven1969,Elliott1974} is a powerful tool for the calculation of properties of random systems, from the electronic structure of alloys \cite{Soven1969,Kirkpatrick1970} 
 to the elasticity of composite materials \cite{Budiansky1965,Hill1965,Gubernatis1975}.  
 It has been used with great success in the study of rigidity percolation problem \cite{Gennes1976,Feng1985,Garboczi1985} and, more recently, in the dynamics of off-lattice systems near a rigidity threshold~\cite{Wyart2010}. 
 It also provides a quantitative theory of the static and dynamic phonon response in the nearly isostatic square lattice in which next-nearest-neighbor springs ($NNN$) are randomly added to a lattice in which all nearest-neighbor-site connected by equivalent springs \cite{Mao2010}. In this lattice, the effective-medium $NNN$ spring constant $\kappa_m(\Prob,\omega)$ has a zero-frequency  limit, $\kappa_m(\Prob,0)$, that scales as $\Prob^2 \sim (\Delta z)^2$ leading to lengths and frequencies that scale as $l^*\sim (\Delta z)^{-1}$ and $\omega^* \sim (\Delta z)^{1}$ as they do in the jamming problem. The finite-frequency spring constant is a scaling function of $\omega/\omega^*$ at small $\Prob$: $\kappa_m(\Prob,\omega)= \kappa_m(\Prob,0) h(\omega/\omega^*)$. As $\Prob$ is increased, there is a crossover from nonaffine response at small $\Prob$ to affine response with $\kappa_m(\Prob,0) = \Prob \kappa$ at larger $\Prob$, much as there is in networks of semi-flexible polymers as the average polymer length is increased \cite{Head2005,Wilhelm2003}

\subsection{Review of Results}

In this paper, we use the CPA to explore in detail the static and dynamic properties of the kagome lattice as it is moved away from isostaticity by the random addition of $NNN$ springs. Our approach is identical to that employed in our treatment of the square lattice \cite{Mao2010}, but calculations are considerably more complicated because each unit cell has three sites rather than a single one. The results of our calculations follow what we expect from our experience with the homogeneous square and kagome lattices and with the random square lattice. The bulk and shear moduli are nonzero and proportional to the $NN$ spring constant $k$ when $\Prob \rightarrow 0$.  The static $NNN$ effective medium spring constant $\kappa_m(\Prob,0)$ scales as $\Prob^2$ leading to $l^* \sim (\Delta z)^{-1}$ and $\omega^* \sim (\Delta z)$ in agreement with cutting arguments.
As is the case in the square lattice, $\kappa_m(\Prob,\omega)/\kappa_m(\Prob,0)$ is a basically a scaling function of $\omega/\omega^*$ at small $\Prob$, but with small yet important deviations at small $\omega$ that describe Rayleigh scattering, i.e., a mean-free-path that scales as $\omega^{-3}$ in $2d$.

The outline of this paper is as follows.  In Sec.~\ref{SEC:KagoElas} we review the elasticity of the homogeneous nearly isostatic kagome lattice.  In Sec.~\ref{SEC:CPA} we discuss the CPA on random nearly isotropic kagome lattice with the $NNN$ bonds randomly occupied with probability $\Prob$.  In Sec.~\ref{SEC:RESULTS} we discuss the results of the CPA calculation, including the crossover of $\fcNNNm$ from $\Prob^2$ to $\Prob$ behavior as  $\Prob$ increase or $\fcNNN$ decreases, and the rapid increase of scattering at the characteristic frequency $\freq_{D}^{*}\sim\Delta z$.

\begin{figure}
\centerline{\includegraphics[width=.4\textwidth]{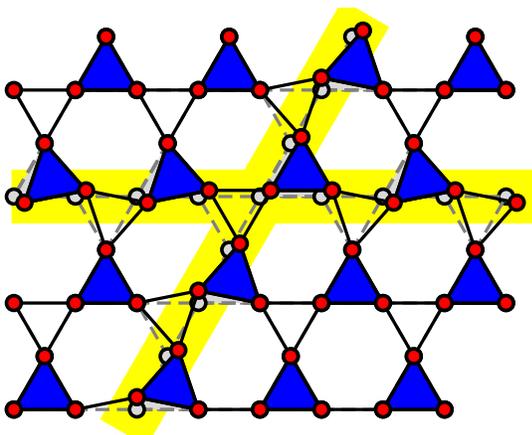}}
\caption{The kagome lattice and its floppy modes, with the reference state in gray and deformed state in red.  Two of its floppy modes are shown in this figure marked by the yellow ribbons.}
\label{FIG:kagome}
\end{figure}


\section{Homogeneous nearly isostatic kagome lattice and its elasticity}
\label{SEC:KagoElas}

\subsection{Expansion of elastic energy in general lattice models}
In this section we briefly review the elastic energy in central-force network models, in which the elastic energy $\EE$ can be written as a sum of the energy of each central-force bond
\begin{eqnarray}
	\EE = \sum_{b} \EE_b (R_b) ,
\end{eqnarray}
where $R_b$ is the length of the bond and $U_b$ is the potential energy of the bond as a function of the length.  We consider a displacement field on the network that maps particle $\ell$ which is at position $\RefrPosil$ to a new position $\TargPosil=\RefrPosil+\ul$, thus the length of bond $b$ between particle $\ell$ and $\ell'$ is changed into
\begin{eqnarray}
	R_b = \vert \TargPosilp-\TargPosil \vert .
\end{eqnarray}
We refer to the original space in which particle $\ell$ is at $\RefrPosil$ as the \emph{reference space}, and the space after applying the displacement field as the \emph{target space}.  We consider harmonic potentials
\begin{eqnarray}
	\EE_b = \frac{k_b}{2} (R_b-R_{bR})^2 ,
\end{eqnarray}
where $R_{bR}$ is the rest length of the bond, and $k_b$ is the spring constant.  The length $R_b$ can be expanded for small displacement $\mathbf{u}$ as
\begin{eqnarray}
	R_b &=& R_{b0} + \uv _{b0} \cdot \ub \nonumber\\
	&&  + \frac{1}{2R_{b0}} \ub \cdot \big(\mathbf{I}-\uv _{b0}\uv _{b0}\big) \cdot \ub +O(\ub^3)  ,
\end{eqnarray}
where $R_{b0} = \vert \RefrPosilp-\RefrPosil \vert$, $\ub=\ulp-\ul$, and $\uv _{b0}=(\RefrPosilp-\RefrPosil)/\vert \RefrPosilp-\RefrPosil \vert$ is the unit vector pointing along the bond in the reference space.  Thus we have
\begin{eqnarray}
	\EE_b &=& \frac{k_b}{2}(R_{b0}-R_{bR})^2 + f_b \, \uv _{b0} \cdot \ub \nonumber\\
	&& + \frac{1}{2} \ub \cdot \Big\lbrack
		k_b \uv _{b0}\uv _{b0} +\frac{f_b}{R_{b0}} \big(\mathbf{I}-\uv _{b0}\uv _{b0}\big)
	\Big\rbrack\cdot \ub ,
\end{eqnarray}
where $f_b=U_b'(R)=k_b(R_{b0}-R_{bR})$ is the magnitude of the force on the bond in the reference space.  In general we consider the case in which the reference state is in mechanical equilibrium, which means that the total force on each particle vanishes
\begin{eqnarray}
	\mathbf{f}_{\ell} = \sum_{b(\ell,\ell')} f_{b} \uv _{b0} = 0
\end{eqnarray}
where the sum $\sum_{b(\ell,\ell')}$ is over all occupied bonds connected to $\ell$.  However, to capture the properties of random networks, which often carry residual stress, the length of each bond is not necessarily at its rest length, i.e., $R_{b0}-R_{bR}\ne 0$ in general.

The change of the elastic energy from the reference space to the target space of the whole system is then a quadratic form of the displacement field
\begin{eqnarray}\label{EQ:DeltVUUGene}
	\Delta \EE = \sum_{b}
	 \frac{1}{2} \ub \cdot \Big\lbrack
		k_b \uv _{b0}\uv _{b0} +\frac{f_b}{R_{b0}} \big(\mathbf{I}-\uv _{b0}\uv _{b0}\big)
	\Big\rbrack\cdot \ub ,
\end{eqnarray}
which can also been written as
\begin{eqnarray}
	\Delta \EE = \sum_{b}
	 \frac{1}{2} \Big\lbrack k_b \big(\ub^{\parallel}\big)^2 +\frac{f_b}{R_{b0}} \big(\ub^{\perp}\big)^2
	\Big\rbrack ,
\end{eqnarray}
where $\ub^{\parallel}$ is the component of $\ub$ parallel to $\uv _{b0}$ and $\ub^{\perp}$ is the component perpendicular to $\uv _{b0}$.

By doing a gradient expansion on the displacement field,
\begin{eqnarray}
	\ub = R_{b0} \, e_{b0k} \, \partial_{k} \mathbf{u}(\mathbf{r}),
\end{eqnarray}
where $\mathbf{u}(\mathbf{r})$ is the displacement field at position $\mathbf{r}$,
we recover the elastic energy of the continuum theory
\begin{eqnarray}
	\Delta \EE = \int d\mathbf{r} K_{ijkl} \partial_{k} u_i \partial_{l} u_j ,
\end{eqnarray}
with
\begin{eqnarray}
	K_{ijkl} &=& \sum_{\mathbf{b}}\frac{R_{b0}^2}{2v_0}  e_{b0k}e_{b0l} \nonumber\\
	&& \cdot\Big\lbrack
		k_b e_{b0i}e_{b0j} +\frac{f_b}{R_{b0}} \big(\delta_{ij}-e_{b0i}e_{b0j}\big)
	\Big\rbrack
\end{eqnarray}
where the summation $\sum_{\mathbf{b}}$ is over bonds connecting to one particle, and we are using a simple lattice with one particle per unit cell in this illustration.  The volume of a unit cell is denoted by $v_0$.

\subsection{Elastic energy of the kagome lattice}
The kagome lattice is a lattice with three particles per unit cell, and we shall use the following displacement vector to denote the deformation of the lattice
\begin{eqnarray}\label{EQ:BasisSix}
	\vu_{\ell} = (u_{\ell,1,x},u_{\ell,1,y},u_{\ell,2,x},u_{\ell,2,y},u_{\ell,3,x},u_{\ell,3,y}),
\end{eqnarray}
where $\ell$ labels the unit cell and $(1,2,3)$ label the particles in the unit cell as in Fig.~\ref{FIG:kagome}.  As we have discussed, the elastic energy can be expanded in small displacement field $\vu$, and to leading order we have the quadratic form
\begin{eqnarray}\label{EQ:DefiDM}
	\Delta \EE = \frac{1}{2} \sum_{\ell,\ell'} \vu_{\ell} \cdot \DM_{\ell,\ell'} \cdot \vu_{\ell'} ,
\end{eqnarray}
which can be built from the analysis as in Eq.~(\ref{EQ:DeltVUUGene}), but generalized to the case of the kagome lattice which has three sites per unit cell.  The matrix $\DM$ is called the dynamical matrix of the lattice.  This elastic energy can be written in momentum space as
\begin{eqnarray}
	\Delta \EE = \frac{1}{2N^2}\sum_{\vq,\vq'} \vu_{\vq} \cdot \DM_{-\vq,\vq'} \cdot \vu_{\vq'} ,
\end{eqnarray}
where the Fourier transform into momentum space is defined as
\begin{eqnarray}\label{EQ:FTDefi}
	\vu_{\vq} &=& \sum_{\ell} \vu_{\ell} e^{-i \vq \cdot \RefrPosil} \nonumber\\
	\vu_{\ell} &=& \frac{1}{N} \sum_{\vq} \vu_{\vq} e^{i \vq \cdot \RefrPosil} ,
\end{eqnarray}
where $N$ is the number of unit cells.
The dynamical matrix for the homogeneous kagome lattice with all $NN$ bonds occupied with springs of spring constant $\fcNN$ and all $NNN$ bonds with springs of spring constant $\fcNNN$ is a $6\times 6$ matrix given by
\begin{eqnarray}\label{EQ:DMCons}
	\DM_{\vq,\vq'} &=& N\delta_{\vq,\vq'} \DM_{\vq} (\fcNN,\fcNNN) \nonumber\\
	\DM_{\vq}(\fcNN,\fcNNN)  &=& \fcNN \sum_{m\in NN} \vB^{NN}_{m,\vq} \vB^{NN}_{m,-\vq} \nonumber\\
	&& + \fcNNN \sum_{m\in NNN} \vB^{NNN}_{m,\vq} \vB^{NNN}_{m,-\vq} ,
\end{eqnarray}
where the $\vB$ vectors and their derivation are given in App.~\ref{APP:DM}.

\subsection{The homogeneous kagome lattice and its low energy theory}\label{SEC:HKL}
There are six translational degrees of freedom per unit cell in the kagome lattice giving rise to six phonon branches.  Of these, three are optical branches with frequencies of order $\sqrt{k}$, two are acoustic branches with sound velocities of order $\sqrt{k}$, and one is the anomalous branch.  The later three branches, which determines the low-energy elastic theory of the kagome lattice, have modes in the space spanned by the following three vectors
\begin{eqnarray}
	\nu_1 &=& (1/\sqrt{3})( 1,0,1,0,1,0 ) \nonumber\\
	\nu_2 &=& (1/\sqrt{3})( 0,1,0,1,0,1) \nonumber\\
	\nu_3 &=& \big( -\frac{1}{\sqrt{3}},0,\frac{1}{2\sqrt{3}},-\frac{1}{2},\frac{1}{2\sqrt{3}},\frac{1}{2} \big) ,
\end{eqnarray}
which correspond to two translations of the whole unit cell in $x$ and $y$ directions and the rotation of the unit-cell triangle around its center.  The low-energy theory is governed by the $3\times 3$ reduced dynamical matrix obtained by integrating out the three high-energy optical branches, as shown in App.~\ref{APP:RDM}.

For small momentum, $\vert\mathbf{q}\vert<q_H^*=4 \sqrt{3 \kappa/k}$, the reduced dynamical matrix is simply diagonalized by longitudinal and transverse acoustic phonons (which are linear combinations of $\nu_1$ and $\nu_2$) with speeds of sound $c_{L}=\sqrt{3\fcNN}/4$ and $c_T=\sqrt{\fcNN}/4$ and the rotational mode with a characteristic frequency $\freqst_{O}=\sqrt{6\fcNNN}$ at $\vq=0$.  The bulk modulus $B$ and the shear modulus $G$ are related, respectively, to the longitudinal and transverse sound velocities through 
\begin{equation}
	c_L^2 = (B+ G)/\varrho; \qquad c_T^2 = G/\varrho , 
\end{equation}
where $\varrho$ is the mass density, which because there are three atoms per unit cell, is equal to $3$ in our units.  Thus, $B=3k/8$ and $G= 3k/16$.  
There is only weak mixing between the rotational modes and the acoustic phonons, and the system is isotropic.  

\begin{figure}
\includegraphics{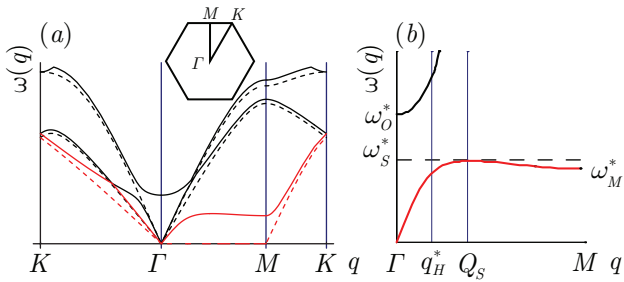}
\caption{(color online) (a) Phonon dispersion along symmetry
directions. The dotted lines are for $\kappa = 0$ and the solid lines
are for $\kappa = 0.02$. The floppy and anomalous branches
are in red. (b) shows anomalous and shear modes along $\Gamma M$
and indicates characteristic frequencies and wavenumbers.  Frequencies $\freqst_{O}$, $\freqst_{S}$ and $\freqst_{M}$ are defined in the text. (From Ref.~\cite{Souslov2009})}
\label{fig:kagome-modes}
\end{figure}

For large momentum $\vert\mathbf{q}\vert>q_H^*=4 \sqrt{3 \kappa/k}$, strong mixing between the transverse acoustic modes and the rotational modes occurs, and the strong anisotropy of the isostatic state is retrieved.  The mixing is maximal along $q_x=0$ and symmetry equivalent directions, which we refer to as the isostatic directions, and the resulting two modes are shown in Fig.~\ref{fig:kagome-modes}.  The anomalous branch, with frequency of order $\fcNNN$, is the lower branch of the two. In the limit of $\fcNNN=0$, the lattice becomes isostatic, the isotropic region is squeezed to the origin, and the anomalous modes reduce to the isostatic floppy modes with zero frequency along $q_x=0$ ($\Gamma M$ line in Fig.~\ref{fig:kagome-modes}) and symmetry equivalent directions as depicted in Fig.~\ref{FIG:kagome}.
The name ``anomalous modes'' follows the nomenclature of Ref.~\cite{Wyart2005a}, referring to the modes developed from the floppy modes as the system is moved away from the isostatic point.  For a more detailed discussion of the low energy theory of the elasticity of the kagome lattice, see App.~\ref{APP:RDM}.

\begin{figure}
\centerline{\includegraphics[width=.35\textwidth]{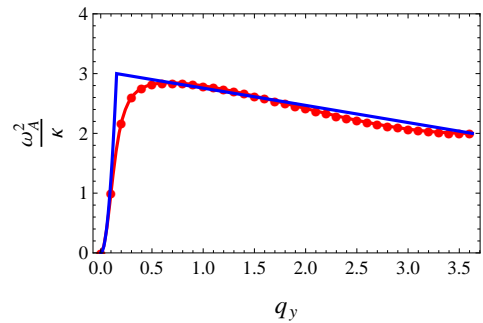}}
\caption{(color online) Eigenvalue of the isostatic mode along isostatic directions, e.g., $q_x=0$, for $\fcNNN=5\times 10^{-4}$.  The eigenvalue of the full $6\times 6$ dynamical matrix, $\freq^2$, normalized by $\fcNNN$, is denoted by the red dots, and the eigenvalue of the $3 \times 3$ reduced dynamical matrix~(\ref{EQ:DMSPLF}) is denoted by the red line.  The blue line represent the approximation~(\ref{EQ:omegaA}) we used in the asymptotic calculation in the $\fFunc$ in CPA. }
\label{FIG:IsosMode}
\end{figure}

Of particular interest is the frequency of the anomalous modes in the vicinity of $q_x=0$.  The squared frequency of these modes can be written as
\begin{equation}
\omega^2 ( \vq ) = \omega_A^2 ( q_y) + c_x^2 q_x^2 ,
\end{equation}
where $c_x = c_L = \sqrt{3k}/4$. The function $\omega_A^2(q_y)$ is plotted in Fig.~\ref{FIG:IsosMode}.  It reaches a maximum value of $(\freqst_S)^2$ at a $2d$ saddle point at $q_y = Q_S$ and a local minimum value of $(\freqst_M)^2$ at the zone edge $q_y = Q_M = 2 \pi/\sqrt{3}$.  For small $\kappa$, $Q_s \simeq 4(3 \kappa/2k)^{1/4}$, $(\freqst_S)^2 \simeq 3\kappa $ and $(\freqst_M)^2 \simeq 2 \kappa$.
All of the characteristic frequencies $\freqst_O>\omega_S>\omega_M$ are proportional to $\sqrt{\kappa}$ for small $\kappa$. $\omega_A^2(q_y)$ is well approximated between $q_y = Q_S$ and $q_y= Q_M$ by
\begin{equation}\label{EQ:omegaA}
\omega_A^2 \approx \frac{1}{Q_M-Q_S}[Q_M \omega_S^2 - Q_S \omega_M^2 - q_y (\omega_S^2 - \omega_M^2)] ,
\end{equation}
as is evident from Fig.~\ref{FIG:IsosMode}. This relation will prove useful in our evaluation of integrals in our CPA analysis in Sec.~\ref{SEC:CPA}

Lengths scaling as $\kappa^{-1/2}$ can be extracted from the phonon dispersion relations in various ways.  One length is the hybridization length $l_H^*$ obtained from the hybridization wavenumber $q_H^*=4 \sqrt{3 \kappa/k} = l_H^{*-1}$ separating the domain of predominantly transverse phonon behavior at low $q_y$ from the domain of predominantly rotation behavior at high $q_y$.  Other lengths can be obtained by comparing the $c_x q_x^2$ term in $\omega^2(\vq)$  to $\omega_M^2$ and $\omega_S^2$: $q_M^*=\omega_M/c_x = l_m^{*-1}=(8/\sqrt{6})\sqrt{\kappa/k}$ and $q_S^*=\omega_S/c_x = l_S^{*-1}=4\sqrt{\kappa/k}$. An interesting property of $\omega_A^2 (q_y)$ is that the hybridization frequency $\omega_H^*$ obtained by setting $q_y=q_H^*$ in the transverse phonon frequency is identical to $\omega_S^*$: $c_T q_H^* \equiv \omega_H^* = \omega_S^*$.

One experimentally relevant quantity is the Fourier transform of the finite temperature static phonon correlation function
$\mathscr{G}_{\mu,\nu}(\lv, \lv')$:
\begin{equation}
 \mathscr{G}_{\mu,\nu} (\vq) = k_B T \sum_{\alpha} \frac{e_{\mu}^{\alpha}(-\vq) e_{\nu}^{\alpha} (\vq)}{\omega_{\alpha}^2(\vq)} ,
\end{equation}
where $\mu$ and $\nu$ label the basis defined in Eq.~(\ref{EQ:BasisSix}) of the $6-$dimensional space of $\vu$, $\alpha$ labels the phonon band, and $e_{\mu}^{\alpha}(\vq)$ is the $6$-dimensional eigenvector associated with mode $(\alpha, \vq)$.  This correlation function is in the zero-frequency limit and thus independent of dissipation of the system.  The quantities $\omega_{\alpha}^2(\vq)$ are merely the eigenvalues of the dynamical matrix with the zero-frequency value of the spring constant (the effective medium spring constant can depend on frequency as we discuss below in the CPA).  The diverging length scale $\lst\sim\kappa^{-1/2}$ can be extracted this way from the static phonon correlation function in experiments.

\section{The Coherent Potential Approximation on the random nearly isostatic kagome lattice}\label{SEC:CPA}
The CPA is a widely used method in the study of disordered systems~\cite{Soven1969,Feng1985,Garboczi1985}.  In it, a random system is mapped into an effective medium with no disorder that is described by a Green's function with a suitable self-energy that can capture the effect of the disorder average of the randomness.  To achieve this, one imposes a self-consistency constraint that the effective medium Green's function perturbed by the presence of single impurity in the effective medium reduces to the effective medium Green's function when averaged over the probability distribution of the impurity. More specifically, the $T$-matrix of this perturbation vanishes upon averaging over configurations that contain and do not contain the impurity.

For the case of the nearly isostatic kagome lattice, the effective medium has all $NNN$ bonds occupied with an effective-medium spring of spring constant $\fcNNNm (\Prob,\omega)$, and the effective medium Green's function is identical to that of a homogeneous system with $\kappa = \fcNNNm(\Prob,\omega)$. The CPA procedure consists of replacing one arbitrary $NNN$ bond with a new bond of spring constant $\fcNNNs$, which takes on the value $\fcNNN$ with probability $\Prob$ (bond occupied) and the value $0$ with probability $1-\Prob$ (bond unoccupied). This procedure leads to a modified dynamical matrix
\begin{eqnarray}
	\DM^{V} = \DM + \PPV ,
\end{eqnarray}
where
\begin{eqnarray}
	\PPV_{\vq,\vq'}(\fcNN,\fcNNN)  = (\fcNNNs-\fcNNNm) \vB^{NNN}_{1,\vq} \vB^{NNN}_{1,-\vq'} ,
\end{eqnarray}
where $1$ represents the arbitrary $NNN$ bond we have chosen to replace into $\fcNNNs$. This form of $\PPV$ follows directly from the calculations leading to Eq.~(\ref{EQ:DMCons}).   It depends on the wavenumbers $\vq$ and $\vq'$ because the perturbed system is not translationally invariant.

The phonon Green's function for the effective medium is
\begin{eqnarray}
	\GF_{\vq}(\freq)=\big\lbrack \freq^2 \mathbf{I} -\DM_{\vq}  \big\rbrack ^{-1} .
\end{eqnarray}
In the perturbed system with one bond replaced, the Green's function becomes
\begin{eqnarray}
	\GF_{\vq,\vq'}^{V}(\freq)=\big\lbrack \freq^2 \mathbf{I} -\DM^{V}  \big\rbrack ^{-1}_{\vq,\vq'}
\end{eqnarray}
and is no longer translationally invariant.
This Green's function can be expanded for small $\PPV$
\begin{eqnarray}
	\!\! \GF_{\vq,\vq'}^{V}\!\! &=& \!\!(\mathbf{I}-\GF\cdot\PPV)^{-1}_{\vq,\vq'}\cdot\GF_{\vq'} \nonumber\\
	&\simeq& \!\! N\delta_{\vq,\vq'}\GF_{\vq} + \GF_{\vq}\!\!\cdot\!\PPV_{\vq,\vq'}\!\!\cdot\!\GF_{\vq'} \nonumber\\
	&& \!\!+ \frac{1}{N}\sum_{\vq_1}\! \GF_{\vq}\!\!\cdot\!\PPV_{\vq,\vq_1}\!\!\cdot\!\GF_{\vq_1}\!\!\cdot\!\PPV_{\vq_1,\vq'}\!\!\cdot\!\GF_{\vq'}\! +\!\cdots  ,\!
\end{eqnarray}
where we have dropped the frequency $\freq$ dependence which is the same for every $\GF$ and $\PPV$.  This series can be written as
\begin{eqnarray}
	\GF_{\vq,\vq'}^{V}
	= N\delta_{\vq,\vq'}\GF_{\vq} + \GF_{\vq}\cdot\TM_{\vq,\vq'}\cdot\GF_{\vq'} ,
\end{eqnarray}
where
\begin{eqnarray}
	\!\!\!\TM_{\vq,\vq'} \!\!&\equiv&\!\! \PPV_{\vq,\vq'} + \frac{1}{N} \sum_{\vq_1}\! \PPV_{\vq,\vq_1}\!\!\cdot\!\GF_{\vq_1}\!\!\cdot\!\PPV_{\vq_1,\vq'} \nonumber\\
	&&\!\! + \frac{1}{N^2}\sum_{\vq_1,\vq_2}\! \PPV_{\vq,\vq_1}\!\!\cdot\!\GF_{\vq_1}\!\!\cdot\!\PPV_{\vq_1,\vq_2}\!\!\cdot\!\GF_{\vq_2}\!\!\cdot\!\PPV_{\vq_2,\vq'} \!
	\nonumber\\
	&& + \cdots ,
\label{eq:T-matrix1}
\end{eqnarray}
is the $T$-matrix expressed in the wavenumber basis.

In the CPA, the effective medium spring constant $\fcNNNm$ is determined by requiring that the average value of $\GF_{\vq,\vq'}^{V}$ be equal to $N\delta_{\vq,\vq'}\GF_{\vq}$ or equivalently that the disorder average of the $T$-matrix vanish:
\begin{eqnarray}\label{EQ:SCET}
	\Prob \TM \vert_{\fcNNNs=\fcNNN} +(1-\Prob) \TM\vert_{\fcNNNs=0} =0 .
\end{eqnarray}
The evaluation of the $T$-matrix is simplified by the following identity,
\begin{align}
	&\frac{1}{N}\sum_{\vq_1}\! \PPV_{\vq,\vq_1}\!\!\cdot\!\GF_{\vq_1}\!\!\cdot\!\PPV_{\vq_1,\vq'}\\
	&= (\fcNNNs-\fcNNNm)^2 \vB^{NNN}_{1,\vq}\nonumber\\
	&\quad\quad\times\frac{1}{N}\!\left(\sum_{\vq_1}\vB^{NNN}_{1,-\vq_1} \cdot \GF_{\vq_1}\cdot\vB^{NNN}_{1,\vq_1}\!\right)\! \vB^{NNN}_{1,-\vq'} \\
& = -(\fcNNNs-\fcNNNm) \PPV_{\vq,\vq'} \fFunc (\fcNNNm, \freq) ,
\end{align}
where
\begin{eqnarray}\label{EQ:fDef}
	\fFunc(\fcNNNm, \freq) = -v_0 \int_{1BZ} \frac{d^{2}\mathbf{q}}{4\pi^2}\, \vB^{NNN}_{1,-\vq} \cdot \GF_{\vq}(\freq) \cdot \vB^{NNN}_{1,\vq} ,
\end{eqnarray}
with $v_0 = \sqrt{3}/2$ the area of the unit cell in real space and $4\pi^2 /v_0= 8\pi^2/\sqrt{3}$ is the area of the first Brillouin zone in reciprocal space.  The integral is over the first Brillouin zone.  The Green's function $\GF_{\vq}(\freq)$ is the phonon Green's function in the effective medium so it depends on $\fcNNNm$. Using these relations in Eq.~(\ref{eq:T-matrix1}) gives
\begin{eqnarray}
	\TM_{\vq,\vq'} = \frac{\PPV_{\vq,\vq'}}{1+(\fcNNNs-\fcNNNm)\fFunc (\fcNNNm, \freq) }.
\end{eqnarray}
Thus, the self-consistency equation~(\ref{EQ:SCET}) requires that
\begin{eqnarray}\label{EQ:CPASCE}
	\fFunc(\fcNNNm, \freq) \fcNNNm^2 - (1+\fcNNN \fFunc(\fcNNNm, \freq)) \fcNNNm + \Prob \fcNNN =0 ,
\end{eqnarray}
from which one can solve for the effective medium $NNN$ spring constant $\fcNNNm$ for any given $\Prob$ and $\freq$. The form of this solution at small $\fcNNNm$ depends on the behavior of the function $\fFunc(\fcNNNm, \freq)$ at small $\fcNNNm$, which is in turn determined by the form of the anomalous mode along the $q_x=0$ and other isostatic directions.  Details of the  calculation of $\fFunc(\fcNNNm, \freq)$ are presented in App.~\ref{APP:fInt}.

In the following discussion unless otherwise stated, we use reduced units with $\fcNN=1$ and lattice constant $a=1$, and thus unitless spring constants, and elastic moduli, and frequencies: $\fcNNN/\fcNN\to\fcNNN$, $Ga^2/\fcNN\to G$, and $\freq/\sqrt{\fcNN}\to\freq$.

\section{Results and Discussion}\label{SEC:RESULTS}
\subsection{CPA solution at zero frequency: static response}
\begin{figure}
\centerline{\includegraphics[width=.4\textwidth]{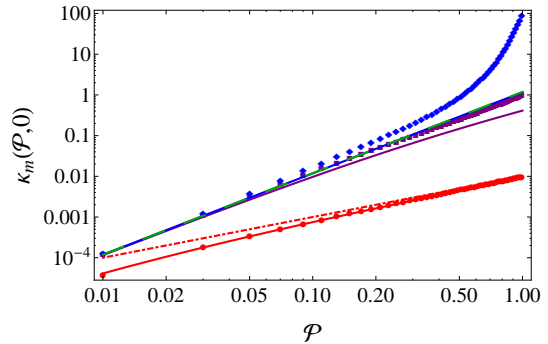}}
\caption{(color online) CPA solution at zero frequency.  Data points show the numerical solution $\fcNNNm$ as a function of $\Prob$ at $\freq=0$ of the CPA self-consistency equation~(\ref{EQ:CPASCE}) with the full $6\times 6$ dynamical matrix.  $NNN$ bond spring constant $\fcNNN = 10^{-2}, 10^{0}, 10^{2}$ are shown in red dots, purple squares, and blue diamonds respectively.  The analytical solution~(\ref{EQ:SCESoluSP}) of these three cases at small $\Prob$ are shown in red, purple, and blue lines.
Also shown are the nonaffine ($\fcNNNm=A\Prob^2$) and affine ($\fcNNNm=\Prob\fcNNN$ at $\fcNNN=10^{-2}$) limits in green dashed line and red dash-dotted line.  At large $\Prob$ the numerical solution, especially the one for $\fcNNN = 10^{2}$ deviate significantly from the nonaffine limit form because Eq.~(\ref{EQ:SCESP}) is an approximation at small $\Prob$ by ignoring the highest order term in Eq.~(\ref{EQ:CPASCE}).}
\label{FIG:ZeroFreqCPA}
\end{figure}

We first consider the case of $\freq=0$, which characterizes the static response of the system.  For small $\Prob$, we expect that the effective medium spring constant $\fcNNNm$ also to be small and that we can, therefore, ignore the $\fFunc(\fcNNNm, \freq) \fcNNNm^2$ term in the CPA self-consistency equation~(\ref{EQ:CPASCE}).  Using the asymptotic small $\kappa_m$ limit  $\fFunc(\fcNNNm, 0)=\bb/\sqrt{\kappa_m}$, where $\bb= 5(1-\sqrt{2/3})$, derived in App.~\ref{APP:fInt} [Eq.~(\ref{EQ:fzerofreq})], we obtain the equation for $\fcNNNm$ at small $\Prob$
\begin{eqnarray}\label{EQ:SCESP}
	\fcNNNm + \bb \fcNNN \sqrt{\fcNNNm} -\Prob \fcNNN =0 ,
\end{eqnarray}
which has the solution
\begin{eqnarray}\label{EQ:SCESoluSP}
	\fcNNNm(\Prob,0) = \Big\lbrack \frac{-\bb \fcNNN + \sqrt{\bb^2 \fcNNN^2 + 4\Prob \fcNNN}}{2}\, \Big\rbrack^2.
\end{eqnarray}
This solution has two limits:
\begin{equation}\label{EQ:AffiNona}
	\fcNNNm(\Prob,0) \simeq
	\begin{cases}
		\Aa \Prob^2  & \text{if
    	$\Prob\ll (\bb^2/4)\fcNNN$  ,}
    	\\
   	\Prob\fcNNN & \text{if $\Prob\gg (\bb^2/4)\fcNNN$  ,} 
	\end{cases}
\end{equation}
where $\Aa= 1/\bb^2 =3(5+2\sqrt{6})/25$.
In the first case, $\fcNNN \sqrt{\fcNNNm} \gg \fcNNNm$, and the solution for $\fcNNNm$ is obtained by ignoring the first term in Eq.~(\ref{EQ:SCESP}); in the second case, the opposite is true, and $\fcNNNm$ is obtained by ignoring the second term in this equation.  In the second case, every $NNN$ bond distorts in the same way under stress, and response is affine.  In the first case $\fcNNNm=A\Prob^2 \ll \Prob\fcNNN $, indicating that the response is nonaffine with local rearrangements in response to stress.  Within the CPA, this result emerges because of the divergent elastic response encoded in $\GF$ (and $\fFunc(\fcNNNm,0)$) as $\fcNNNm\to 0$ (See App.~\ref{APP:fInt}). The nonaffine regime arises when $NNN$ springs are strong enough for the second term in Eq.~(\ref{EQ:SCESP}) to dominate the first. As $\fcNNN$ approaches zero at fixed $\Prob$, distortions produced by the extra bond decrease and the nonaffine regime becomes vanishingly small.  Numerical solutions of the CPA self-consistency equation~(\ref{EQ:CPASCE}) with the full $6\times 6$ dynamical matrix is plotted in Fig.~\ref{FIG:ZeroFreqCPA}, along with a comparison to the analytical solution~(\ref{EQ:SCESoluSP}) and the two asymptotic forms in Eq.~(\ref{EQ:AffiNona}).

This crossover of the effective medium spring constant $\fcNNNm$ between $\Prob^2$ and $\Prob$ is different from the nonaffine-affine crossover in the case of random nearly isostatic square lattice~\cite{Mao2010} which describes the response of the square lattice to shear stress, because the shear modulus (more precisely, $C_{44}$) is proportional to the effective medium spring constant $\fcNNNm$ for the $NNN$ bonds in the square lattice.  In the case of kagome lattice in this Paper, the shear modulus is finite and proportional to $\fcNN$, whereas $\fcNNNm$ mainly determines the rigidity of the lattice with respect to the floppy mode, which are essentially rotations of the triangles as shown in Fig.~\ref{FIG:kagome}.

Length and frequency scales can be extracted in the static limit much as they were extracted in the homogeneous case discussed in Sec.~\ref{SEC:KagoElas}.  The finite temperature static phonon correlation function $\mathscr{G}$ is the inverse of the dynamical matrix evaluated at $\omega = 0$, whose eigenvalues and eigenvectors are identical to those of the homogeneous case with $\kappa$ replaced by $\kappa_m \equiv \kappa_m(\Prob, 0)$.  The eigenvalues allows us to identify frequencies by taking the square roots of the appropriate eigenvalues of $\DM$:
\begin{equation}
	\omega^*_O=\sqrt{6 \kappa_m} >
	\omega^*_S=\sqrt{3\kappa_m} >\omega^*_M =\sqrt{2 \kappa_m}.
\label{eq:omega*}
\end{equation}
Unlike the situation homogeneous lattices, these frequencies are not equal to any physical dynamical-mode frequency of the system.  They do, however, provide information about the static properties of the phonon correlation function $\mathscr{G}$ that could in principle be measured at finite temperature via scattering or particle tracking experiment. They also lead to diverging lengths just as they do in the homogeneous case:
\begin{equation}
l^* \equiv l_H^* = \frac{1}{\sqrt{3}} \, l_S^*= \sqrt{2} \,l_M^*=\frac{1}{4 \sqrt{3 \kappa_m}}= \frac{1}{\sqrt{3 \Aa}}\frac{1}{\Delta z}
\end{equation}

\subsection{CPA solution at finite frequency: dynamic response and damping}\label{SEC:CPAFF}
\begin{figure}
\centerline{\includegraphics[width=.4\textwidth]{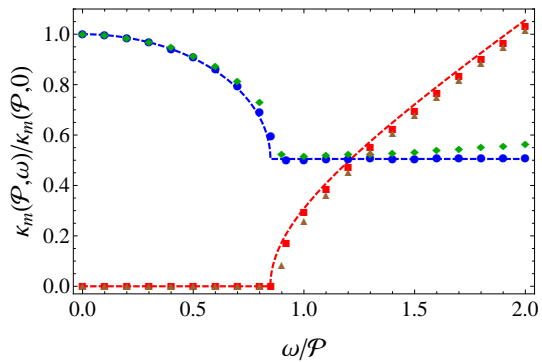}}
\caption{(color online) CPA solution at finite frequency for $\Prob=0.01$ and $\Prob=0.05$.  The numerical solution to Eq.~(\ref{EQ:CPASCE}) with the full $6\times 6$ dynamical matrix is shown as the data points.  Blue circles and red squares represent real and (negative of) imaginary parts of $\fcNNNm$ at $\Prob=0.01$, and green diamonds and brown triangles represent the real and (negative of) imaginary parts of $\fcNNNm$ at $\Prob=0.05$.  The asymptotic form~(\ref{EQ:CPAASYMSOLU}) is shown as the blue (real) and red (negative of imaginary) lines.  In this plot frequency is rescaled by $\Prob$, and the effective medium spring constant $\fcNNNm$ is rescaled by its value at zero frequency which is real.
}
\label{FIG:FiniFreqCPA}
\end{figure}

For finite frequency $\freq$, the effective medium spring constant is complex, $\fcNNNm(\Prob,\freq)=\fcNNNm'(\Prob,\omega) -i\fcNNNm''(\Prob,\omega)$, where the imaginary part $\fcNNNm''(\Prob,\omega)$, which describes damping of phonons in this random network, is odd in $\freq$ and positive for $\freq>0$.  From the analysis for the static limit $\freq=0$, we see that the interesting case is the nonaffine regime with $\Prob\ll (\bb^2/4)\fcNNN$, in which the self-consistency equation~(\ref{EQ:SCESP}) simplifies to
\begin{eqnarray}\label{EQ:CPALO}
	\fFunc(\fcNNNm,\freq) \fcNNNm = \Prob .
\end{eqnarray}
In the static limit, $\fFunc(\fcNNNm,0) \sim\fcNNNm^{-1/2}$ is singular in the $\fcNNNm\rightarrow 0$ limit. As we show in App.~\ref{APP:fInt}, at finite frequency, 
$\fFunc(\fcNNNm,\freq) \sim [\sqrt{(3 \fcNNNm - \omega^2)/\fcNNNm} - \sqrt{(2 \fcNNNm - \omega^2)/\fcNNNm}]/\fcNNNm$, 
which leads to
\begin{eqnarray}\label{EQ:CPAASYMSOLU}
	\fcNNNm(\Prob,\freq) =  \frac{3\Prob^2}{25} \Bigg(
		5 +2\sqrt{6}\sqrt{1-\frac{25\freq^2}{18\Prob^2}}
	\Bigg) ,
\end{eqnarray}
as depicted in Fig.~\ref{FIG:FiniFreqCPA}.  Taking $\freq=0$, this solution reduces to the zero-frequency solution [Eq.~(\ref{EQ:AffiNona})] in the nonaffine limit.  Equation~(\ref{EQ:CPAASYMSOLU}) develops an imaginary part when $\vert\freq\vert >2\sqrt{3}\Prob/5$, which, as we discussed above must be negative for $\freq>0$.  It is straightforward to see that this solution satisfies the scaling form $\fcNNNm(\Prob,\freq)=\fcNNNm(\Prob,0) h(\freq/\freq^*)$, as does the CPA effective $NNN$ spring constant in the square lattice Ref.~\cite{Mao2010}.
This solution shows a rapid increase of damping beyond a characteristic frequency
\begin{eqnarray}
	\freqst_{D}=\frac{2\sqrt{3}\Prob}{5},
\end{eqnarray}
marking another characteristic frequency that scales also as $\Prob$.

Numerical solution of the CPA self-consistency equation~(\ref{EQ:CPASCE}) using the full $6\times 6$ dynamical matrix is also shown in Fig.~\ref{FIG:FiniFreqCPA}.  We see that the asymptotic form~(\ref{EQ:CPAASYMSOLU}) captures the solution fairly well.

This special behavior of the imaginary part of the effective medium spring constant $\fcNNNm$ is related to the phonon spectrum of the kagome lattice.  As we have discussed in Sec~\ref{SEC:KagoElas}, at low frequencies, there is only very weak mixing between the rotational branch which is strongly affected by the $NNN$ bonds and the acoustic phonon branches which are only very weakly scattered by the $NNN$ bonds, and thus the damping to the acoustic phonons is very weak.  On the other hand, for higher frequencies the transverse phonon strongly mixes with the rotational modes, making the damping rapidly increase with scatterings from the $NNN$ bonds beyond $\freqst_D$, although the exact value of $\freqst_D$ may be an artifact of the CPA method.  
The weak scattering below $\freqst_{D}$ is not captured by the asymptotic form~(\ref{EQ:CPAASYMSOLU}) for small $\fcNNNm$ because the function $\fcNNNm(\Prob,\freq)$ in Eq.~(\ref{EQ:CPAASYMSOLU}) was obtained using the dominant small $\fcNNNm$ limit of the integral $\fFunc(\fcNNNm,\freq)$.  There are, however, contributions to this integral that do not diverge and that contribute a subdominant imaginary part to $\fcNNNm$, even when $\freq < \freq_D^*$, that
is of order $\Prob^3 \freq^2$ at small $\freq$ corresponding to Rayleigh scattering.  More discussion is included in App.~\ref{APP:fInt}.

In the homogeneous case, the eigenvalues of the dynamical matrix lead naturally to the identification of characteristic frequencies $\omega_S^*$, and $\omega_M^*$ that vanish as $\sqrt{\kappa}$ in the limit of $\kappa \rightarrow 0$. In the random case, we have to deal with both the frequency-dependence $\kappa_m(\Prob, \omega)$ and the fact that it is a complex number, and we must ask whether these frequencies have any real meaning.  As discussed earlier in Sec.~\ref{SEC:HKL}, we can extract frequencies from the static dynamical matrix in exactly the same way that we did for the homogeneous case, and they satisfy
\begin{equation}
	\omega_O^* = 3.85 \omega_D^* > 
	\omega_S^* = 2.72 \omega_D^* > \omega_M^*= 2.22 \omega_D^* > \omega_D^* .
\end{equation}
Thus all of these frequencies are greater than the frequency $\omega_D$.  
As a result, the signatures in the phonon dispersion relation including the hybridization and the saddle point are washed out by the strong scattering, as is shown in Fig.~\ref{FIG:Scattering}.

\subsection{Phonon density of states}

\begin{figure}
	\centering
	 \subfigure[]{\includegraphics[width=.35\textwidth]{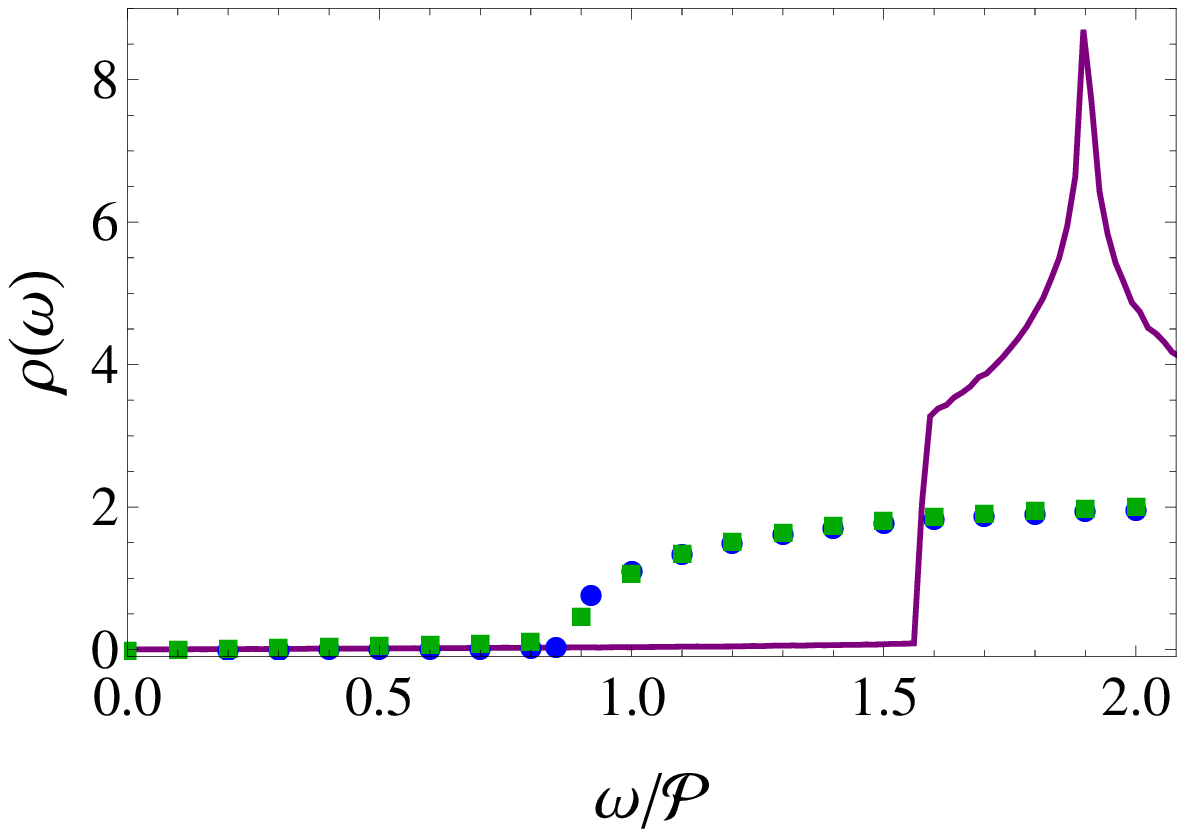}}
	 \subfigure[]{\includegraphics[width=.38\textwidth]{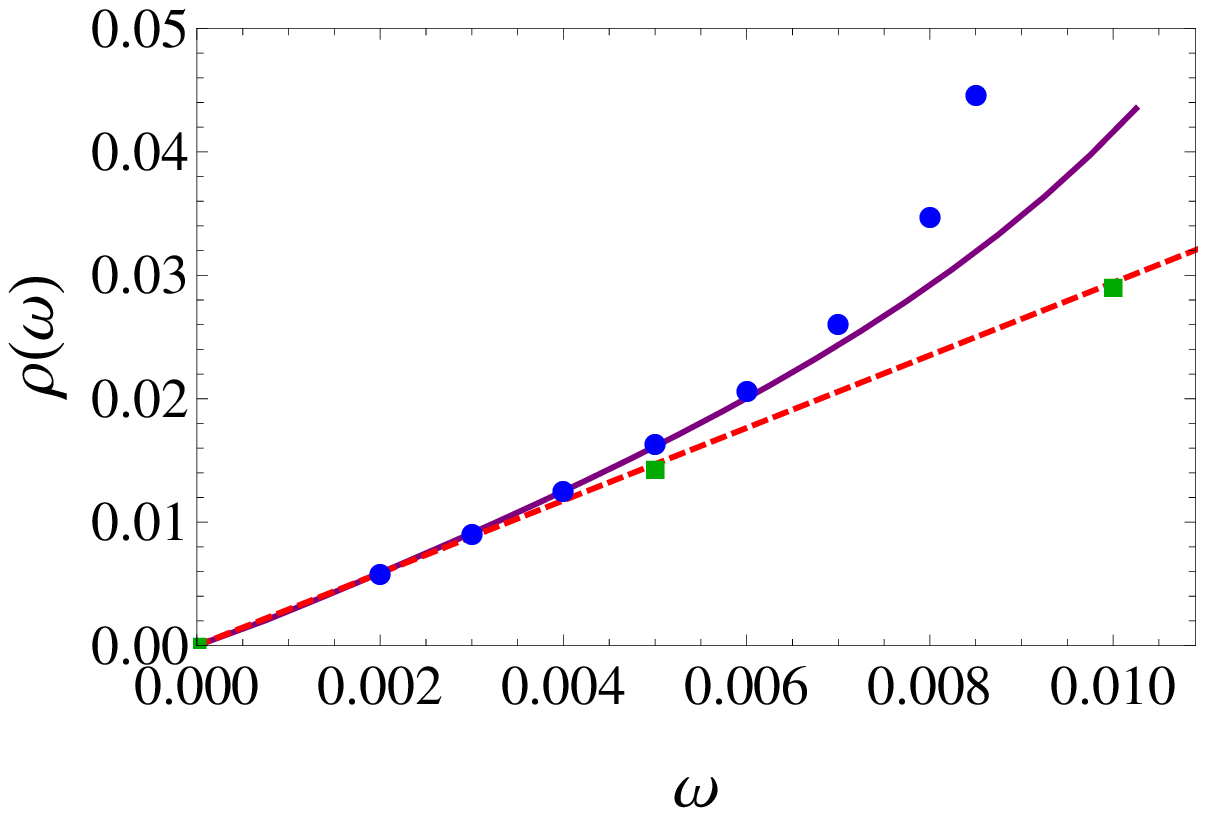}}
		\caption{(color online) (a) The phonon DOS at $\Prob=0.01$ (blue circles) and $\Prob=0.05$ (green squares) of the CPA effective medium and the pure kagome lattice with the $NNN$ spring constant equal to the zero frequency effective medium value $\fcNNNm(\Prob,\freq)$ for $\Prob=0.01$ (purple line).  The frequency is rescaled by $\Prob$.  (b) The phonon DOS at small frequency for CPA effective medium (color scheme the same as in (a)).  The Debye DOS defined in Eq.~(\ref{EQ:DebyDOS}) is also shown as the red dashed line.
		}
\label{FIG:DOS}
\end{figure}

The phonon density of states (DOS) can be calculated from the retarded Green's function through
\begin{eqnarray}
	\DOS(\freq) = -\frac{1}{\pi}  \textrm{Tr}\,\textrm{Im} \GF (\vq,\freq)
\end{eqnarray}
where the trace is over both momentum $\vq$ and the phonon modes.  Using this we get the phonon DOS of the effective medium solved from the CPA, as plotted in Fig.~\ref{FIG:DOS}.  As a comparison, we also show the phonon DOS of a periodic kagome lattice with the $NNN$ spring constant equal to $\fcNNNm(\Prob,0)$, which is real valued.

For small frequencies, at which the imaginary part of the CPA solution $\fcNNNm(\Prob,\freq)$ is very small, the two DOS are very close, and can be fitted nicely by the Debye-like total DOS of the transverse and the longitudinal phonons
\begin{eqnarray}\label{EQ:DebyDOS}
	\DOS_{s}(\freq) = \frac{\freq}{(4\pi/\sqrt{3})c_L^2}+\frac{\freq}{(4\pi/\sqrt{3})c_T^2}
\end{eqnarray}
where $c_L^2=3\fcNN/16$ and $c_T^2=\fcNN/16$ are respectively the longitudinal and transverse speed of sound (we have taken $\fcNN=1$ as stated earlier).

At the critical frequency $\freqst_D$, the imaginary part of $\fcNNNm(\Prob,\freq)$ rapidly increase, inducing a rapid increase of the phonon DOS. On the other hand, the periodic lattice exhibit a jump in DOS at $\freqst_M=\sqrt{2\fcNNNm(\Prob,0)}$ corresponding to the minimum of the phonon dispersion relations at the edge of the 1BZ.  At $\freqst_S=\sqrt{3\fcNNNm(\Prob,0)}$ the DOS of the periodic lattice has a logarithmic singularity, corresponding to the saddle point of the phonon dispersion at $Q_S \simeq 4(3\fcNNNm/2\fcNN)^{1/4}$ on the isostatic directions~\cite{Souslov2009}.  For the CPA effective medium, this singularity is totally washed out due to the strong damping beyond $\freqst_D$, which is similar to the case of the square lattice~\cite{Mao2010}.

\subsection{Phonon scattering and the Ioffe-Regel limit}
From the CPA solution at finite frequency, we identified a frequency scale $\freqst_D$ beyond which phonon scattering rapidly increase.  In this subsection we examine the scattering of phonons in more detail.

\begin{figure}
\centerline{\includegraphics[width=.45\textwidth]{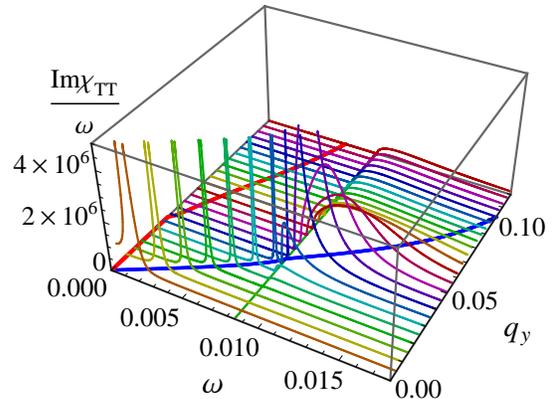}}
\caption{(color online) Scattering of phonons characterized using the imaginary part of the transverse component of the phonon response function $\textrm{Im}\chi_{TT}(\vq,\freq)$ as a function of $\freq$ for various values of $q_y$ (we took $q_x=0$ to follow the isostatic $\Gamma M$ direction).  The green line in the bottom plane marks $\freqst_D$, the blue and red lines marks the solution $\freq'$ and $\freq''$ of the equation $\freq^2 - \freq_{A}(\vq,\fcNNNm(\Prob,\freq))^2=0$, which solves for the pole of the Green's function for the anomalous branch.  The derivation of $\freq_{A}$ is shown in App.~\ref{APP:RDM}.}
\label{FIG:Scattering}
\end{figure}

The scattering of the transverse phonons is characterized by the imaginary part of the phonon response function projected to the transverse direction $\textrm{Im}\chi_{TT}(\vq,\freq)$.  The phonon response function is defined as
\begin{eqnarray}
	\mathbf{\chi}_{\mu,\nu}(\ell,t;\ell',t')\equiv \frac{\delta  u_{\mu}(\ell,t)}{\delta F_{\nu}(\ell',t')},
\end{eqnarray}
where $t$ and $t'$ label time, $\mu$ and $\nu$ label the basis defined in Eq.~(\ref{EQ:BasisSix}) of the $6-$dimensional space of $\vu$.
This response function is related to the phonon Green's function through $\mathbf{\chi}=-\GF$.  The imaginary part of the transverse component of this response function $\textrm{Im}\chi_{TT}(\vq,\freq)$ characterizes the scattering of the transverse phonon by disorder.  $\textrm{Im}\chi_{TT}(\vq,\freq)$ is calculated for small momentum and frequency using the asymptotic CPA solution~(\ref{EQ:CPAASYMSOLU}), and shown in Fig.~\ref{FIG:Scattering}.  Also shown in the figure is the frequency at which the phonon Green's function of the anomalous branch has a complex pole, which is solved from the equation $\freq^2 - \freq_{A}(\vq,\fcNNNm(\Prob,\freq))^2=0$, which characterizes the dynamic dispersion relation.  We use the form of $\freq_{A}$ as defined in App.~\ref{APP:RDM} for this calculation.  It is clearly shown in the figure that below $\freqst_D$, the response function has Dirac-delta peaks at the frequencies determined by the transverse phonon dispersion relation $\freq=c_T q_y$.  Above $\freqst_D$ the imaginary part $\freq''$ takes off, and the phonon peaks progressively broaden, showing that the transverse phonon is no longer a good eigenstate of the system.  Furthermore, the hybridization of the transverse phonon and the rotational modes into the anomalous mode and the van Hove singularities in the density of states are washed out by the strong scattering.  As a result, $\freqst_O$, $\freqst_S$, $\freqst_M$ no longer play meaningful role in the dynamic response function.

The strength of the scattering can be characterized by the Ioffe-Regel (IR) limit, which states that the plane wave states are no longer well defined if the mean free path $l_{\text{mfp}}$ is comparable to or less than the phonon wavelength $\lambda$.  An equivalent condition is that the relaxation time becomes comparable to the period of the wave, i.e., $\freq''\sim \freq'$.  The solution for the positions of the complex poles of the Green's function of the anomalous branch $\freq^2 - \freq_{A}(\vq,\fcNNNm(\Prob,\freq))^2=0$ shows that the imaginary part $\freq''$ becomes comparable to the real part $\freq'$ not far beyond $\freqst_D$ and that $\omega^*_{\text{IR}}\sim \freqst_D \sim\Delta z$. The associated IR length scale can be derived from $\freqst_D$ and $c_T$ to be of order $l_{\text{IR}}\sim \sqrt{\fcNN/\fcNNNm} \sim \Delta z^{-1}$.

This IR length scale differs from that in jammed solids, $l_d \sim \Delta z^{-1/2}$ as derived in Refs.~\cite{Xu2009,Vitelli2010} (called $l_s$ in Ref.~\cite{Wyart2010}).  This discrepancy can be attributed to the different scaling of the shear modulus $G$.  In the kagome lattice, the shear modulus $G$ is proportional to $\fcNN$ and thus scales as $\Delta z^0$, whereas in jammed solids $G\sim\Delta z$.  Thus, the transverse speeds of sound scales as $(\Delta z)^{0}$ and $(\Delta z)^{1/2}$ in these two cases respectively.  In both cases, the frequency beyond which plane wave states are strongly scattered is $\freqst\sim\Delta z$.  Therefore, the scattering length scale, $c_T/\freqst$ are respectively $\lst\sim(\Delta z)^{-1}$ and $l_d\sim(\Delta z)^{-1/2}$ in the kagome lattice and jammed solids.

\subsection{Comparison between different random nearly isostatic systems}
Up to now, three examples of random nearly isostatic systems have been studied, including the random nearly isostatic square lattice, random nearly isostatic kagome lattice discussed in this Paper, and jammed solids near point J.  In all cases, the characteristic frequency for the onset of the anomalous mode plateau $\freqst\sim\Delta z$ and the isostatic length scale $\lst\sim (\Delta z)^{-1}$.  On the other hand, the scaling of elastic moduli in the three cases are different because of different \emph{network architecture}: in the square lattice $G\sim\fcNNNm\sim(\Delta z)^2$ and $B\sim\fcNN\sim (\Delta z)^{0}$, in kagome lattice $G,B\sim \fcNN\sim(\Delta z)^{0}$, and in jammed solids $G\sim\Delta z$ and $B\sim\fcNN\sim (\Delta z)^{0}$ (a jump from zero to finite value at point J).  As a result, the scattering length scales are also different.  The square lattice is anisotropic, and we studied the scattering of the $u_x$ vibrations along $q_y$ direction and found that the scattering length corresponded to the point $M$ in the first Brillouin zone, $q_y=\pi$, or $l_{x,\text{IR}}=a$.  In kagome lattice $l_{\text{IR}}\sim c_T/\freqst\sim(\Delta z)^{-1}$.  In jammed solids, the IR length scale $l_d\sim c_T/\freqst\sim(\Delta z)^{-1/2}$.

In Ref.~\cite{Wyart2010}, Wyart studied transport properties 
of amorphous solids modeled by an isostropic random network near its percolative rigidity threshold. In this system, both $B$ and $G$ vanish as $\Delta z$, and the crossover frequency between plane-wave and strongly scattered states is $\omega^* \sim\Delta z$.  Both the  longitudinal and transverse sound velocities scale as $(\Delta z)^{1/2}$, and the IR length $l_{\text{IR}}\sim c_{L,T}/\omega^*$ scales as $(\Delta z)^{-1/2}$ for both modes. The CPA self-consistency equation for low frequency in Ref.~\cite{Wyart2010}, Eq.~(7), can be rewritten in the form
\begin{eqnarray}
	k_M^2 - (\Delta z) k_m +\freq^2 =0,
\end{eqnarray}
by ignoring the $\freq^3$ term which correspond to Rayleigh scattering in the second subequation.  The solution to this equation
\begin{eqnarray}
	k_M = \frac{\Delta z}{2} \Big( 1 + \sqrt{1 - \frac{4\freq^2}{\Delta z^2}}  \Big),
\end{eqnarray}
has a form very similar to the that of the effective medium $NNN$ spring constant $\fcNNNm$ in the kagome lattice as shown in Eq.~(\ref{EQ:CPAASYMSOLU}).  Although $k_M(\omega=0) \sim\Delta z$ in the amorphous solid and $\kappa_M(\omega=0)\sim (\Delta z)^2$ in the kagome lattice scale differently with $\Delta z$, the frequency dependence of $k_M(\omega)/k_M(\omega=0)$ and $\kappa_M(\omega)/\kappa_M(\omega=0)$ are almost identical: they are both of the form $a+b \sqrt{1-(\omega/\omega_D^*)^2}$ where $a$ and $b$ are constants and $\omega_D \sim \Delta z$. The $\freq^3$ term ignored in the above analysis leads to Rayleigh scattering and corresponds to the subdominant terms in $\fFunc$ discussed in Sec.~\ref{SEC:CPAFF}.  The difference between $\freq^3$ in Ref.~\cite{Wyart2010} and $\freq^2$ for the kagome lattice is the spatial dimension.

To summarize, we examined the random nearly isostatic kagome lattice via the CPA, we obtained effective-medium $NNN$ spring constant $\fcNNNm$ that scales with the occupancy probability $\Prob \sim \Delta z$ of the $NNN$ bonds as $\Prob^2$ at small $\Prob$.  Below the characteristic frequency $\freqst_D\sim\Prob$, there is only weak damping of acoustic phonons arising from Rayleigh scattering, whereas above $\freqst_D$ scattering increases rapidly and the system shows proximity to the IR limit.  We compare the kagome lattice to other nearly isostatic systems including the square lattice, jammed solids near point J, and a model random isotropic network~\cite{Wyart2010}.  The characteristic frequency scale $\freqst\sim\Delta z$, marking both the onset of the plateau of the anomalous modes and the strong scattering of plain wave states, is found to be a universal property of all of these systems.  
The elastic modulus $G,B$ and thus the transport length scale depends on the \emph{network architecture} and are not universal.

\noindent
{\it Acknowledgments\/}---%
This work was supported in part by NSF-DMR-0804900.

\appendix

\section{The dynamical matrix of the kagome lattice}\label{APP:DM}
To construct the dynamical matrix of the kagome lattice, we use the form of the elastic energy given in Eq.~(\ref{EQ:DeltVUUGene}).  Because we consider the reference state of all bonds at their rest length, we have $f_b=0$, thus there is only projection of $\Disp$ onto the direction along the bond.  We first consider the case of simple lattice with one particle in each unit cell and rewrite Eq.~(\ref{EQ:DeltVUUGene}) as
\begin{eqnarray}
	\Delta \EE &=& \sum_{b}
	 	\frac{k_b}{2} \big\lbrack
	 		(\ulone-\ultwo) \cdot \uv_{\ell_1 \ell_2}
	 	\big\rbrack ^2  \nonumber\\
	 &=& \sum_{\ell,\ell'} \sum_{b} \frac{k_b}{2}
	 	\ul \cdot \uv_{\ell_1 \ell_2}  (\delta_{\ell,\ell_1}-\delta_{\ell,\ell_2})
	 	\nonumber\\
	 	&& \quad\quad\times(\delta_{\ell',\ell_1}-\delta_{\ell',\ell_2})
	 	\uv_{\ell_1 \ell_2}\cdot\ulp ,
\end{eqnarray}
where $\ell_1 , \ell_2$ labels the two particles connected by the bond $b$.  Thus the dynamical matrix $\DM$, as defined in Eq.~(\ref{EQ:DefiDM}), is given by
\begin{eqnarray}
	\DM_{\ell,\ell'}&=&\sum_{b} k_b \uv_{\ell_1 \ell_2}  (\delta_{\ell,\ell_1}-\delta_{\ell,\ell_2})
	 	\nonumber\\
	 	&& \quad\quad\times(\delta_{\ell',\ell_1}-\delta_{\ell',\ell_2})\uv_{\ell_1 \ell_2}.
\end{eqnarray}
It is convenient to express the dynamical matrix in momentum space via the Fourier transform defined in Eq.~(\ref{EQ:FTDefi})
\begin{eqnarray}
	\DM_{\vq,\vq'} &=& \sum_{\ell,\ell'} e^{-i\vq\cdot \rv_{\ell}+i\vq'\cdot \rv_{\ell'}}\DM_{\ell,\ell'} \nonumber\\
	&=& \sum_{\ell,\ell'} e^{-i\vq\cdot \rv_{\ell}+i\vq'\cdot \rv_{\ell'}}\sum_{\ell_1}\sum'_{\ell_2} k_b \uv_{\ell_1 \ell_2}  \nonumber\\
	&&		\,\times (\delta_{\ell,\ell_1}-\delta_{\ell,\ell_2})(\delta_{\ell',\ell_1}-\delta_{\ell',\ell_2}) \uv_{\ell_1 \ell_2} \nonumber\\
	&=& N \delta_{\vq,\vq'} \sum_{\bv} k_{\bv} (1-e^{-i\vq\cdot\bv})\nonumber\\
	 	&& \quad\quad\times(1-e^{i\vq\cdot\bv}) \uv_{\bv}\uv_{\bv}
\end{eqnarray}
where the $'$ above the summation of $\ell_2$ denote a summation over particles connected to $\ell_1$, and
$\bv=\rv_{\ell'}-\rv_{\ell}$ represent the bonds connected to an arbitrary particle (note the difference from $b$ in the previous equation, which represent all bonds in the system).  One can define the dynamical matrix for translational invariant system as
\begin{eqnarray}
	\DM_{\vq,\vq'} &=& N\delta_{\vq,\vq'} \DM_{\vq}  \nonumber\\
	\DM_{\vq} &=& \sum_{\bv} k_{\bv} (1-e^{-i\vq\cdot\bv})(1-e^{i\vq\cdot\bv}) \uv_{\bv}\uv_{\bv} \nonumber\\
	&=& \sum_{m} k_{m} \vB_{m,\vq}\vB_{m,-\vq},
\end{eqnarray}
where the summation $m$ is over bonds connected to an arbitrary particle, and the vector
\begin{eqnarray}
	\vB_{m,\vq} = (1-e^{-i\vq\cdot\bv_m})\uv_{\bv_m}
\end{eqnarray}
is a convenient way to express the dynamical matrix.

For the kagome lattice, which has three particles per unit cell, one need to modify the above construction of the dynamical matrix, and in the basis of
\begin{eqnarray}
	\vu_{\ell} = (u_{\ell,1,x},u_{\ell,1,y},u_{\ell,2,x},u_{\ell,2,y},u_{\ell,3,x},u_{\ell,3,y}),
\end{eqnarray}
with particles $1,2,3$ labeled as in Fig.~\ref{FIG:kagome}, the dynamical matrix can be expressed as
\begin{eqnarray}
	\DM_{\vq,\vq'} &=& N\delta_{\vq,\vq'} \DM_{\vq} (\fcNN,\fcNNN) \nonumber\\
	\DM_{\vq}(\fcNN,\fcNNN)  &=& \fcNN \sum_{m\in NN} \vB^{NN}_{m,\vq} \vB^{NN}_{m,-\vq} \nonumber\\
	&& + \fcNNN \sum_{m\in NNN} \vB^{NNN}_{m,\vq} \vB^{NNN}_{m,-\vq} ,
\end{eqnarray}
with the $\vB$ vectors for $NN$  bonds for each unit cell (each bond is counted once)
\begin{eqnarray}
	\vB^{NN}_{1,\vq} &=& \Big( -\frac{1}{2},-\frac{\sqrt{3}}{2},\frac{1}{2},\frac{\sqrt{3}}{2},0,0 \Big) \nonumber\\
	\vB^{NN}_{2,\vq} &=& \Big( 0,0,1,0,-1,0 \Big) \nonumber\\
	\vB^{NN}_{3,\vq} &=& \Big( \frac{1}{2},-\frac{\sqrt{3}}{2},0,0,-\frac{1}{2},\frac{\sqrt{3}}{2} \Big) \nonumber\\
	\vB^{NN}_{4,\vq} &=& \Big( -\frac{1}{2},-\frac{\sqrt{3}}{2},\frac{1}{2} e^{-i\big(\frac{1}{2}q_x+\frac{\sqrt{3}}{2}q_y\big)}, \nonumber\\
	&& \quad		 \frac{\sqrt{3}}{2}e^{-i\big(\frac{1}{2}q_x+\frac{\sqrt{3}}{2}q_y\big)},0,0 \Big) \nonumber\\
	\vB^{NN}_{5,\vq} &=& \Big( 0,0,-e^{-iq_x},0,1,0 \Big)  \nonumber\\
	\vB^{NN}_{6,\vq} &=& \Big( \frac{1}{2},-\frac{\sqrt{3}}{2},0,0,
			-\frac{1}{2} e^{-i\big(-\frac{1}{2}q_x+\frac{\sqrt{3}}{2}q_y\big)}, \nonumber\\
	&& \quad		 \frac{\sqrt{3}}{2}e^{-i\big(-\frac{1}{2}q_x+\frac{\sqrt{3}}{2}q_y\big)} \Big),
\end{eqnarray}
and the $\vB$ vectors for $NNN$ bonds for each unit cell
\begin{eqnarray}
	\vB^{NNN}_{1,\vq} &=& \Big( \frac{\sqrt{3}}{2}e^{-iq_x},\frac{1}{2}e^{-iq_x},0,0,-\frac{\sqrt{3}}{2},-\frac{1}{2} \Big) \nonumber\\
	\vB^{NNN}_{2,\vq} &=& \Big( 0,0,0,e^{-i\big(\frac{1}{2}q_x+\frac{\sqrt{3}}{2}q_y\big)},0 ,-1\Big) \nonumber\\
	\vB^{NNN}_{3,\vq} &=& \Big( \frac{\sqrt{3}}{2},\frac{1}{2},0,0,-\frac{\sqrt{3}}{2}e^{-i\big(\frac{1}{2}q_x+\frac{\sqrt{3}}{2}q_y\big)}, \nonumber\\
			&& \quad	 -\frac{1}{2}e^{-i\big(\frac{1}{2}q_x+\frac{\sqrt{3}}{2}q_y\big)} \Big) \nonumber\\
	\vB^{NNN}_{4,\vq} &=& \Big( -\frac{\sqrt{3}}{2},\frac{1}{2},
				 \frac{\sqrt{3}}{2}e^{-i\big(-\frac{1}{2}q_x+\frac{\sqrt{3}}{2}q_y\big)},  \nonumber\\
			&& \quad	 -\frac{1}{2}e^{-i\big(-\frac{1}{2}q_x+\frac{\sqrt{3}}{2}q_y\big)} ,0,0\Big) \nonumber\\
	\vB^{NNN}_{5,\vq} &=& \Big( 0,0,0,1,0,-e^{-i\big(-\frac{1}{2}q_x+\frac{\sqrt{3}}{2}q_y\big)} \Big)  \nonumber\\
	\vB^{NNN}_{6,\vq} &=& \Big( -\frac{\sqrt{3}}{2}e^{iq_x},\frac{1}{2}e^{iq_x},
				\frac{\sqrt{3}}{2},-\frac{1}{2} ,0,0\Big) .
\end{eqnarray}

\section{Calculation of the asymptotic form of the $\fFunc(\fcNNNm,\freq)$ function at small $\fcNNNm$}\label{APP:fInt}
\subsection{The reduced dynamical matrix}\label{APP:RDM}
To calculate the asymptotic form of $\fFunc(\fcNNNm,\freq)$ we first simplify the problem by reduce the dynamical matrix into the space of its three low energy modes by integrating out its three high energy modes~\cite{Souslov2009}.  The resulting low-energy dynamical matrix is conveniently represented in the basis of longitudinal and transverse phonons and the rotational mode (mode $\nu_3$)
\begin{eqnarray}\label{EQ:DMSPLFBS}
	(\nu'_1,\nu'_2,\nu'_3) = (\frac{q_x \nu_1 +q_y \nu_2}{\vert \mathbf{q} \vert},\frac{-q_y \nu_1 +q_x \nu_2}{\vert \mathbf{q} \vert},\nu_3 ),
\end{eqnarray}
in which the dynamical matrix takes the form
\begin{eqnarray}\label{EQ:DMSPLF}
	\tilde{\DM}^{(R)}
	= \fcNN \left(
	\begin{array}{ccc}
	\frac{3q^2}{16} & 0 & \frac{q^2}{16} \cos3\theta \\
	0 & \frac{q^2}{16} & -\frac{q^2}{16}\sin3\theta \\
	\frac{q^2}{16} \cos3\theta & -\frac{q^2}{16}\sin3\theta & \frac{q^2}{16}+\frac{6\fcNNNm}{\fcNN} \\
	\end{array}
	\right) ,
\end{eqnarray}
in leading order of small $\fcNNN$ and quadratic order in $q$ (the cross term of order $\fcNNNm q^2$ is considered higher order and has been dropped).

Eigenmodes of the dynamical matrix are identified by diagonalizing $\tilde{\DM}^{(R)}$.  Strong mixing between the transverse mode and the rotational mode occurs along $q_x=0$ (i.e., $\theta=0$) and symmetry equivalent isostatic directions.  The resulting two eigenvalues (by diagonalizing the lower right $2\times 2$ block) are
\begin{eqnarray}
	\tilde{\freq}_A^2(q) &=& \frac{q^2}{16} + 3 \fcNNNm -\sqrt{\Big(\frac{q^2}{16}\Big)^2 + ( 3 \fcNNNm)^2} \nonumber\\
	\tilde{\freq}_B^2(q) &=& \frac{q^2}{16} + 3 \fcNNNm +\sqrt{\Big(\frac{q^2}{16}\Big)^2 + ( 3 \fcNNNm)^2},
\end{eqnarray}
obtained from the quadratic order of the renormalized $3\times 3$ matrix $\tilde{\DM}^{(R)}$.  The lower eigenvalue $\tilde{\freq}_A^2$ correspond to the \emph{anomalous mode}, which is close to the transverse mode $\nu'_2$ (which is simply $\nu_1$ for $q_x=0$ direction) for $q_y\ll \qst_H = 4\sqrt{3\fcNNNm/\fcNN}$ .  For $q_y\gg \qst_H$ this anomalous mode corresponds to the linear combination of $(\nu'_2-\nu'_3)/2$, which is actually the floppy mode of the kagome lattice in the $\fcNNNm\to 0$ limit, in which $\tilde{\freq}_A\to 0$.

\begin{widetext}
\subsection{Leading order divergence of $\fFunc(\fcNNNm,\freq)$}
The function $\fFunc(\fcNNNm,\freq)$, as given in Eq.~(\ref{EQ:fDef}), can be analyzed using the simplified dynamical matrix~(\ref{EQ:DMSPLF}), which is the leading order form in small $\fcNNNm$ and $q$.  Thus we can obtain an asymptotic analytical calculation of the integral $f$ by projecting from the $6$-dimensional basis in Eq.~(\ref{EQ:BasisSix}) onto the $3$-dimensional basis in Eq.~(\ref{EQ:DMSPLFBS}) built from the three low energy modes of the system
\begin{eqnarray}
	\fFunc(\fcNNNm, \freq) &=& -\int_{1BZ} \frac{d^{2}\mathbf{q}}{8\pi^2/\sqrt{3}} \vB^{NNN}_{1,-\vq} \cdot \GF_{\vq}(\freq) \cdot \vB^{NNN}_{1,\vq} \nonumber\\
	&\simeq& -\int_{1BZ} \frac{d^{2}\mathbf{q}}{8\pi^2/\sqrt{3}} \vB^{NNN}_{1,-\vq} \cdot \Theta^{\textrm{T}}
	 \Theta\cdot  \GF_{\vq}(\freq) \cdot\Theta^{\textrm{T}}\Theta\cdot \vB^{NNN}_{1,\vq} ,
\end{eqnarray}
where
\begin{eqnarray}
	\Theta = \left(
	\begin{array}{cccccc}
	\frac{q_x}{\sqrt{3}q} & \frac{q_y}{\sqrt{3}q} & \frac{q_x}{\sqrt{3}q} & \frac{q_y}{\sqrt{3}q} & \frac{q_x}{\sqrt{3}q} & \frac{q_y}{\sqrt{3}q} \\
	-\frac{q_y}{\sqrt{3}q} & \frac{q_x}{\sqrt{3}q} & -\frac{q_y}{\sqrt{3}q} & \frac{q_x}{\sqrt{3}q} & -\frac{q_y}{\sqrt{3}q} & \frac{q_x}{\sqrt{3}q} \\
	-\frac{1}{\sqrt{3}} & 0 & \frac{1}{2\sqrt{3}} & -\frac{1}{2} & \frac{1}{2\sqrt{3}} & \frac{1}{2} \\
	\end{array}
	\right),
\end{eqnarray}
is the orthogonal transformation from the basis of $\vu_{\ell} = (u_{\ell,1,x},u_{\ell,1,y},u_{\ell,2,x},u_{\ell,2,y},u_{\ell,3,x},u_{\ell,3,y})$ to the basis $(\nu'_1,\nu'_2,\nu'_3)$ in Eq.~(\ref{EQ:DMSPLFBS}) with the longitudinal, transverse, and the rotational mode.  In these new basis, the dynamical matrix is modified by integrating out the high energy modes and keeping to leading order in small $\fcNNNm$ and $q$, which lead to the simple form of Eq.~(\ref{EQ:DMSPLF})~\cite{Souslov2009},
and thus the Green's function can be analyzed correspondingly.  Note that we use the Green's function $\tGF_{\vq}(\freq)$ calculated from the renormalized dynamical matrix~(\ref{EQ:DMSPLF}), so it is different from the bare value $\Theta\cdot  \GF_{\vq}(\freq) \cdot\Theta^{\textrm{T}}$.  
The transformed the $\vB^{NNN}_{1,\vq}$ vector in the basis of $(\nu'_1,\nu'_2,\nu'_3)$ takes the form
\begin{eqnarray}
	\Big( \frac{(e^{-iq_x}-1)(3q_x+\sqrt{3}q_y)}{6\vert \vq\vert},
	\frac{(e^{-iq_x}-1)(\sqrt{3}q_x-3q_y)}{6\vert \vq\vert},
	-\frac{1}{2}(1+e^{-iq_x})\Big) .
\end{eqnarray}

The leading order term of this integral in small $\fcNNNm$ is from $\nu'_3$, the anomalous mode, which has a small frequency of order $\sqrt{\fcNNNm}$ over the whole range of momentum from $q_H$ to the edge of the Brillouin zone along the isostatic directions, and thus correspond to diverging contributions to the $\fFunc$ integral in small $\fcNNNm$.

For an approximation of the $\fFunc$ integral at small $\fcNNNm$, we use the dynamical matrix of the form~(\ref{EQ:DMSPLF}), which kept to leading order in $\fcNNNm$ and quadratic order in $q$.
At small momentum, the dynamical matrix~(\ref{EQ:DMSPLF}) is diagonalized by the basis $(\nu'_1,\nu'_2,\nu'_3)$, and the Green's function $\GF_{\vq}(\freq)$ takes the form of a diagonal matrix
\begin{eqnarray}
	 \tGF_{\vq}(\freq)=\Diag\Big(\frac{1}{\freq^2-\frac{3q^2}{16}}, \frac{1}{\freq^2-\frac{q^2}{16}},\frac{1}{\freq^2-6\fcNNNm-\frac{q^2}{16}}\Big),
\end{eqnarray}
which is isotropic and valid for small momentum $\vert \vq\vert<\qst_H$.
Thus, the small momentum region contribute to $\fFunc$ the following terms
\begin{eqnarray}\label{EQ:fSmall}
	\!\!\!\fFunc_{<} (\fcNNNm,\freq) \!\!&=&\!\!\! -\!\int_{\vert \vq\vert<\qst_H} \frac{dq_x dq_y}{8\pi^2/\sqrt{3}} \Big\lbrace
		\frac{(1-\cos q_x)(3q_x+\sqrt{3}q_y)^2}{18(q_x^2+q_y^2)(\freq^2-\frac{3(q_x^2+q_y^2)}{16})}
		 + \frac{(1-\cos q_x)(\sqrt{3}q_x-3q_y)^2}{18(q_x^2+q_y^2)(\freq^2-\frac{q_x^2+q_y^2}{16})}
		 + \frac{1+\cos q_x}{2(\freq^2\!-6\fcNNN\!-\frac{q_x^2+q_y^2}{16})}
	\Big\rbrace .\,
\end{eqnarray}

At large momentum, the dynamical matrix can be diagonalized to leading order in $\fcNNNm$ in the basis $(\nu'_1,\frac{\nu'_2+\nu'_3}{\sqrt{2}},\frac{\nu'_2-\nu'_3}{\sqrt{2}})$, in which $\vB_{1,\vq}$ takes the form
\begin{eqnarray}
	&& \Big( \frac{(e^{-iq_x}-1)(3q_x+\sqrt{3}q_y)}{6\vert \vq\vert},
	\frac{(e^{-iq_x}-1)(\sqrt{3}q_x-3q_y)}{6\sqrt{2}\vert \vq\vert} - \frac{1}{2\sqrt{2}}(1+e^{-iq_x}) , \nonumber\\
	&& \quad \frac{(e^{-iq_x}-1)(\sqrt{3}q_x-3q_y)}{6\sqrt{2}\vert \vq\vert} + \frac{1}{2\sqrt{2}}(1+e^{-iq_x}) \Big),
\end{eqnarray}
and the Green's function $\GF_{\vq}(\freq)$ takes the form of a diagonal matrix
\begin{eqnarray}\label{EQ:GFLM}
	 \tGF_{\vq}(\freq)=\Diag\Big(\frac{1}{\freq^2-\frac{3(q_x^2+q_y^2)}{16}}, \frac{1}{\freq^2-6\fcNNNm-\frac{q_x^2+q_y^2}{16}},
	\frac{1}{\freq^2-\frac{1}{Q_M-Q_S}[Q_M \omega_S^2 - Q_S \omega_M^2 - q_y (\omega_S^2 - \omega_M^2)] -\frac{3q_x^2}{16}}\Big),
\end{eqnarray}
which is for the direction of $q_x=0$, and we have used the approximated form~\ref{EQ:omegaA} of $\freq_A^2$, that represent the dispersion relation of the anomalous mode at large frequency, as depicted in Fig.~\ref{FIG:IsosMode}.  For this calculation we use the small $\fcNNNm$ values $(\freqst_S)^2 = 3\fcNNNm $ and $(\freqst_M)^2 = 3\fcNNNm $.  

For the other two directions one should change the third term above from $q_x^2$ into the perpendicular direction of the two isostatic directions accordingly.  Thus we need to divide the first Brillouin zone into 3 parts: $\vert\theta-\pi/2\vert<\pi/6$, $\vert\theta-\pi/6\vert<\pi/6$, and $\vert\theta-5\pi/6\vert<\pi/6$, and integrate each of them out separately and then calculate the sum.  Here we just do the $\vert\theta-\pi/2\vert<\pi/6$ part as an example, which uses the form of the Green's function in Eq.~(\ref{EQ:GFLM}).  This part of the integral is
\begin{eqnarray}\label{EQ:fLarge}
	\fFunc_{>,\frac{\pi}{2}} (\fcNNNm,\freq) \!\!&=&\!\! -\frac{2}{8\pi^2/\sqrt{3}} \int_{\qst_H}^{\frac{2\pi}{\sqrt{3}}} dq_y
		\int_{-\frac{\vert q_y\vert}{\sqrt{3}}}^{\frac{\vert q_y\vert}{\sqrt{3}}} dq_x \Big\lbrace
				\frac{(1-\cos q_x)(3q_x+\sqrt{3}q_y)^2}{18(q_x^2+q_y^2)(\freq^2-\frac{3(q_x^2+q_y^2)}{16})} \nonumber\\
		&& \quad 		+ 	\frac{(2+\cos q_x)q_x^2+\sqrt{3}(1-\cos q_x)q_x q_y +3q_y^2}{6(q_x^2+q_y^2)(\freq^2-6\fcNNNm-\frac{q_x^2+q_y^2}{16})} \nonumber\\
		&& \quad
		 + \frac{(2+\cos q_x)q_x^2+\sqrt{3}(1-\cos q_x)q_x q_y +3q_y^2}{6(q_x^2+q_y^2)(\freq^2-\frac{Q_M \omega_S^2 - Q_S \omega_M^2 - q_y (\omega_S^2 - \omega_M^2)}{Q_M-Q_S}-\frac{3q_x^2}{16})}
	\Big\rbrace ,
\end{eqnarray}
and the integral for the other two directions can be calculated similarly.

The leading order contribution to $\fFunc(\fcNNNm,\freq)$ in small $\fcNNNm$ is from the third term in Eq.~(\ref{EQ:fLarge}), which represent the isostatic mode.  We first consider the $\freq=0$ case, for which the leading order term of $\fFunc(\fcNNNm,\freq)$ is
\begin{eqnarray}
	\fFunc_{\frac{\pi}{2},l.o.} (\fcNNNm,0)&=& \frac{2}{8\pi^2/\sqrt{3}} \int_{\qst_H}^{\frac{2\pi}{\sqrt{3}}} dq_y
		\int_{-\frac{\vert q_y\vert}{\sqrt{3}}}^{\frac{\vert q_y\vert}{\sqrt{3}}} dq_x
		\frac{(2+\cos q_x)q_x^2+\sqrt{3}(1-\cos q_x)q_x q_y +3q_y^2}{6(q_x^2+q_y^2)
			\big(\frac{Q_M \omega_S^2 - Q_S \omega_M^2 - q_y (\omega_S^2 - \omega_M^2)}{Q_M-Q_S}+\frac{3q_x^2}{16}\big)} \nonumber\\
	&\simeq& \frac{2}{8\pi^2/\sqrt{3}} \int_{\qst_H}^{\frac{2\pi}{\sqrt{3}}} dq_y
		\frac{(2+\cos q_x)q_x^2+\sqrt{3}(1-\cos q_x)q_x q_y +3q_y^2}{6(q_x^2+q_y^2)} 
	\frac{\pi}{\sqrt{\frac{Q_M \omega_S^2 - Q_S \omega_M^2 - q_y (\omega_S^2 - \omega_M^2)}{Q_M-Q_S}}}
		\frac{4}{\sqrt{3}} \delta(q_x) \nonumber\\
	&\simeq&  \frac{2(1-\sqrt{2/3})}{\sqrt{\fcNNNm}} ,
\end{eqnarray}
where we took the limit of $\fcNNNm\to 0$ and make use of the identity $\lim_{a\to 0} \frac{1}{a^2+x^2} = (\pi/a)\delta(x)$.
Adding up the contribution from $\theta=\pi/6$ and $\theta=5\pi/6$ part we have
\begin{eqnarray}\label{EQ:fzerofreq}
	\fFunc_{l.o.} (\fcNNNm,0)\simeq \frac{5(1-\sqrt{2/3})}{\sqrt{\fcNNNm}}.
\end{eqnarray}
Other terms in Eq.~(\ref{EQ:fSmall}) and (\ref{EQ:fLarge}) contribute higher order terms in small $\fcNNNm$, and is discussed in Sec.~\ref{SEC:HighOrde}.

In the case of $\freq>0$, the leading order term can be calculated in a similar way
\begin{eqnarray}\label{EQ:fzero}
	\fFunc_{\frac{\pi}{2},l.o.} (\fcNNNm,0)&=& -\frac{2}{8\pi^2/\sqrt{3}} \int_{\qst_H}^{\frac{2\pi}{\sqrt{3}}} dq_y
		\int_{-\frac{\vert q_y\vert}{\sqrt{3}}}^{\frac{\vert q_y\vert}{\sqrt{3}}} dq_x
		\frac{(2+\cos q_x)q_x^2+\sqrt{3}(1-\cos q_x)q_x q_y +3q_y^2}{6(q_x^2+q_y^2)
			(\freq^2-\frac{Q_M \omega_S^2 - Q_S \omega_M^2 - q_y (\omega_S^2 - \omega_M^2)}{Q_M-Q_S}-\frac{3q_x^2}{16})} \nonumber\\
	&\simeq& \frac{2}{8\pi^2/\sqrt{3}} \int_{\qst_H}^{\frac{2\pi}{\sqrt{3}}} dq_y
		\, \frac{1}{2} \, \frac{2\pi i (16/3)}{2(4/\sqrt{3})\sqrt{\freq^2-\frac{Q_M \omega_S^2 - Q_S \omega_M^2 - q_y (\omega_S^2 - \omega_M^2)}{Q_M-Q_S}}} \nonumber\\
	&\simeq&  \frac{2}{\sqrt{3\fcNNNm}}\Bigg(\sqrt{3-\frac{\freq^2}{\fcNNNm}}-\sqrt{2-\frac{\freq^2}{\fcNNNm}}\Bigg).
\end{eqnarray}
The $q_x$ integral can either be evaluated using the $\delta$ function trick by assuming an infinitesimal imaginary part of $\freq$ ($\freq\to\freq+i\delta$), or by extending the integral limit of $q_x$ to $(-\infty,\infty)$ (because the integrand decays fast when $q_x$ is large) and using contour integral.  We also assumed that $\vert\sqrt{\frac{Q_M \omega_S^2 - Q_S \omega_M^2 - q_y (\omega_S^2 - \omega_M^2)}{Q_M-Q_S}-\freq^2}\vert\ll 1$ to make the simplification that $\cos q_x \simeq 1$.  Adding up the contribution from $\theta=\pi/6$ and $\theta=5\pi/6$ part we have
\begin{eqnarray}\label{EQ:fffAsym}
	\fFunc_{l.o.} (\fcNNNm,\freq)\simeq \frac{5}{\sqrt{3\fcNNNm}}\Bigg(\sqrt{3-\frac{\freq^2}{\fcNNNm}}-\sqrt{2-\frac{\freq^2}{\fcNNNm}}\Bigg).
\end{eqnarray}
Other terms in Eq.~(\ref{EQ:fSmall}) and (\ref{EQ:fLarge}) contribute higher order terms in small $\fcNNNm$, and is discussed in Sec.~\ref{SEC:HighOrde}.

\subsection{Correction at small frequencies}\label{SEC:HighOrde}
To get the correction to the asymptotic solution of $\fcNNNm(\Prob,\freq)$ as in~(\ref{EQ:fffAsym}), in particular the small imaginary part rather than zero at small frequency, we calculate the imaginary part of $\fFunc$ at small frequencies and solve for the correction to $\fcNNNm(\Prob,\freq)$ perturbatively in the CPA equation.

Because we consider small frequencies $\freq^2<\fcNNNm$, the contribution is from the two acoustic modes, which are isotropic, and thus can be calculated as
\begin{eqnarray}
	\textrm{Im} \fFunc_{L} &\simeq&
	-\frac{2}{8\pi^2/\sqrt{3}} \int _{0}^{\qst_H} dq \int_{0}^{2\pi} d\theta
	\, q\, \textrm{Im} \Big\lbrack
		\frac{q^2 \cos^2\theta (3\cos\theta+\sqrt{3}\sin\theta)^2}{18(\freq^2-\frac{3}{16}q^2+i\delta)}
	\Big\rbrack \nonumber\\
	&\simeq& \frac{20}{27}\freq^2 ,
\end{eqnarray}
and
\begin{eqnarray}
	\textrm{Im} \fFunc_{T} &\simeq&
	-\frac{2}{8\pi^2/\sqrt{3}} \int _{0}^{\qst_H} dq \int_{0}^{2\pi} d\theta
	\, q\, \textrm{Im} \Big\lbrack
		\frac{q^2 \cos^2\theta (\sqrt{3}\cos\theta-3\sin\theta)^2}{18(\freq^2-\frac{1}{16}q^2+i\delta)}
	\Big\rbrack \nonumber\\
	&\simeq& 4\freq^2 .
\end{eqnarray}
Thus we have the correction to Eq.~(\ref{EQ:fffAsym}) that is valid for small $\freq$ as
\begin{eqnarray}
	\fFunc(\fcNNNm,\freq) = \frac{5}{\sqrt{3\fcNNNm}}\Bigg(\sqrt{3-\frac{\freq^2}{\fcNNNm}}-\sqrt{2-\frac{\freq^2}{\fcNNNm}}\Bigg) + i \frac{128}{27} \freq^2.
\end{eqnarray}

We then solve the leading order CPA equation in small $\fcNNNm$ nonaffine regime perturbatively using this corrected form of $\fFunc$ at small $\freq$, and get
\begin{eqnarray}
	\fcNNNm(\Prob,\freq) =  \fcNNNm^{(0)}
	- \frac{256}{135\big(1-\sqrt{2/3}\big)}(\fcNNNm^{(0)})^{3/2}i\freq^2 ,
\end{eqnarray}
where $\fcNNNm^{(0)}$ is the zeroth order solution~(\ref{EQ:CPAASYMSOLU}).  
This correction is very small and can not be observed in our numerical solutions within precision.

\end{widetext}


\end{document}